\documentclass[10pt]{article}
\usepackage{graphicx}
\usepackage{amsmath}
\usepackage{amssymb}
\usepackage{caption2}
\begin{document}
\bibliographystyle{prsty}
\begin{center}
{\large {\bf \sc{  Analysis of  the ${\frac{1}{2}^-}$ and
${\frac{3}{2}^-}$ heavy and doubly heavy
baryon states   with QCD sum rules }}} \\[2mm]
Zhi-Gang Wang \footnote{E-mail,wangzgyiti@yahoo.com.cn.  }     \\
 Department of Physics, North China Electric Power University,
Baoding 071003, P. R. China

\end{center}

\begin{abstract}
In this article, we study the  ${\frac{1}{2}^-}$ and
${\frac{3}{2}^-}$ heavy and doubly heavy baryon states
$\Sigma_Q({\frac{1}{2}^-})$, $\Xi'_Q({\frac{1}{2}^-})$,
$\Omega_Q({\frac{1}{2}^-})$, $\Xi_{QQ}({\frac{1}{2}^-})$,
$\Omega_{QQ}({\frac{1}{2}^-})$, $\Sigma_Q^*({\frac{3}{2}^-})$,
$\Xi_Q^*({\frac{3}{2}^-})$, $\Omega_Q^*({\frac{3}{2}^-})$,
$\Xi^*_{QQ}({\frac{3}{2}^-})$ and $\Omega^*_{QQ}({\frac{3}{2}^-})$
by subtracting the contributions from the corresponding  ${1\over
2}^+$ and ${3\over 2}^+$ heavy and doubly heavy  baryon states with
the QCD sum rules in a  systematic way, and make reasonable
predictions for their masses.
\end{abstract}

 PACS number: 14.20.Lq, 14.20.Mr

 Key words: Heavy baryon states, Doubly heavy baryon states, QCD sum rules

\section{Introduction}

In the past years, there have been several experimental progresses
on the spectroscopy of the heavy baryon states.   In 2006, the Babar
collaboration reported the first observation of the ${3\over 2}^+$
excited singly-charmed baryon state $\Omega_{c}^{*}$  in
 the radiative decay $\Omega_{c}^{*}\rightarrow\Omega_c\gamma$ at the PEP-II
  asymmetric energy $B$-factory at the Stanford linear accelerator center
 \cite{OmegaC}.
By now, the ${1\over 2}^+$ and ${1\over 2}^-$ antitriplet charmed
baryon states ($\Lambda_c^+, \Xi_c^+,\Xi_c^0)$ and
($\Lambda_c^+(2595), \Xi_c^+(2790),\Xi_c^0(2790))$,  and the
${1\over 2}^+$ and ${3\over 2}^+$ sextet charmed baryon states
($\Omega_c,\Sigma_c,\Xi'_c$) and ($\Omega_c^*,\Sigma_c^*,\Xi^*_c$)
have been observed \cite{PDG}. In 2008, the D0 collaboration
reported the first observation of the doubly strange baryon state
$\Omega_{b}^{-}$ in the decay channel $\Omega_b^- \to
J/\psi\thinspace\Omega^-$ with $J/\psi \to \mu^+ \mu^-$ and
$\Omega^- \to \Lambda K^-$  in the $p\bar{p}$ collisions at
$\sqrt{s}=1.96$ TeV  \cite{OmegabD0}. However,  the CDF
collaboration did not confirm the measured  mass
$M_{\Omega_b^-}=(6.165\pm 0.010\thinspace \pm 0.013) \, \rm{GeV}$
\cite{OmegabCDF},  they determined  the mass to be  $(6.0544\pm
0.0068 \pm 0.0009) \,\rm{GeV} $. By now, the $S$-wave bottom baryon
states are far from complete, only the $\Lambda_b$, $\Sigma_b$,
$\Sigma_b^*$, $\Xi_b$, $\Omega_b$ have been observed \cite{PDG}. In
2002,  the SELEX collaboration reported the first observation of a
signal for the doubly charmed baryon state $ \Xi_{cc}^+$ in the
charged decay mode $\Xi_{cc}^+\rightarrow\Lambda_c^+K^-\pi^+$ at the
charm hadro-production experiment (E781) \cite{SELEX2002}, and
confirmed later by the same collaboration in the decay mode
$\Xi_{cc}^+\rightarrow pD^+K^- $ with measured mass $M_{\Xi}=(3518.9
\pm 0.9) \,\rm{ MeV }$ \cite{SELEX2004}. However, the Babar and
Belle collaborations   have not observed any evidence for the doubly
charmed baryon states in $e^{+}e^{-}$ annihilations
\cite{Xi-cc-1,Xi-cc-2}, and did not confirm the SELEX observations.
Moreover, several new excited charmed baryon states, such as the
$\Lambda_c(2765)^+$, $\Lambda_c^+(2880)$, $\Lambda_c^+(2940)$,
$\Sigma_c^+(2800)$, $\Xi_c^+(2980)$, $\Xi_c^+(3080)$,
$\Xi_c^0(2980)$, $\Xi_c^0(3080)$,
  have been observed by the Babar, Belle and CLEO
collaborations in recent years
\cite{ShortRV1,ShortRV2,ShortRV3,BaryonS}, and re-vivified  the
interest in the
  spectroscopy of the charmed  baryon states. On the other hand, the large hadron collider (LHC) will provide
us with the whole excited bottom baryon states \cite{LHC}. Therefore
it is interesting to calculate  the mass spectrum  of the excited
heavy and doubly heavy baryon states with the QCD sum rules.

There have been several approaches to deal with the heavy and doubly
heavy baryon masses, many works focus on the positive-party baryon
states (one can consult Ref.\cite{HH-Roberts} for more literatures);
while the works on the negative-parity baryon states are relatively
few, for example, the non-relativistic quark model
\cite{HH-Roberts,Silvestre-Brac1996}, the relativized  quark model
\cite{Capstick1986,Metsch2006}, the relativistic quark model based
on a quasipotential approach in QCD \cite{HH-Ebert-1,HH-Ebert-2},
the potential approach combined with the QCD sum rules
\cite{HH-Kiselev},  the full QCD sum rules
\cite{HH-Narison,HH-Zhang,Huang0811}, the three-body Faddeev method
\cite{Vijande2007}, etc.

The  QCD sum rules is a powerful theoretical tool in studying the
ground state heavy baryons \cite{SVZ79,PRT85,NarisonBook}. In the
QCD sum rules, the operator product expansion is used to expand the
time-ordered currents into a series of quark and gluon condensates
which parameterize the long distance properties of the QCD. Based on
the quark-hadron duality, we can obtain copious information about
the hadronic parameters at the phenomenological side
\cite{SVZ79,PRT85,NarisonBook}. There have been several works on the
masses of the heavy baryon states with the full QCD sum rules and
the QCD sum rules in the heavy quark effective theory (one can
consult Ref.\cite{Wang0912} for more literatures).

In Ref.\cite{Oka96}, Jido et al introduce a novel approach based on
the QCD sum rules to separate the contributions of   the
negative-parity light flavor  baryons from the positive-parity light
flavor baryons, as the interpolating currents may have non-vanishing
couplings to both the negative- and positive-parity baryons
\cite{Chung82}. Before the work of Jido et al, Bagan et al take the
infinite mass limit for the heavy quarks to separate the
contributions of the positive- and negative-parity heavy baryon
states unambiguously \cite{Bagan93}.

  In Refs.\cite{Wang0912,Wang0101,Wang0101-2,Wang0102}, we follow
Ref.\cite{Oka96} and study the masses and pole residues of the
${\frac{1}{2}}^+$ and ${\frac{3}{2}}^+$ heavy and doubly heavy
baryon states $\Lambda_Q$, $\Xi_Q$, $\Sigma_Q$, $\Xi'_Q$,
$\Omega_Q$, $\Xi_{QQ}$, $\Omega_{QQ}$, $\Sigma_Q^*$, $\Xi_Q^*$,
$\Omega_Q^*$, $\Xi^*_{QQ}$ and $\Omega^*_{QQ}$ in a systematic way
by subtracting the contributions of the corresponding  negative
parity heavy and doubly heavy baryon states. In this article,  we
extend our previous works to study the ${1\over 2}^-$ and ${3\over
2}^-$ heavy and doubly heavy baryon states
$\Sigma_Q({\frac{1}{2}^-})$, $\Xi'_Q({\frac{1}{2}^-})$,
$\Omega_Q({\frac{1}{2}^-})$, $\Xi_{QQ}({\frac{1}{2}^-})$,
$\Omega_{QQ}({\frac{1}{2}^-})$, $\Sigma_Q^*({\frac{3}{2}^-})$,
$\Xi_Q^*({\frac{3}{2}^-})$, $\Omega_Q^*({\frac{3}{2}^-})$,
$\Xi^*_{QQ}({\frac{3}{2}^-})$ and $\Omega^*_{QQ}({\frac{3}{2}^-})$
by subtracting the contributions from the corresponding
${\frac{1}{2}}^+$ and ${\frac{3}{2}}^+$ heavy and doubly heavy
baryon states with the full QCD sum rules.

 The article is arranged as follows:  we derive the
QCD sum rules for the masses and the pole residues of  the negative
parity heavy and doubly heavy baryon states
$\Sigma_Q({\frac{1}{2}^-})$, $\Xi'_Q({\frac{1}{2}^-})$,
$\Omega_Q({\frac{1}{2}^-})$, $\Xi_{QQ}({\frac{1}{2}^-})$,
$\Omega_{QQ}({\frac{1}{2}^-})$, $\Sigma_Q^*({\frac{3}{2}^-})$,
$\Xi_Q^*({\frac{3}{2}^-})$, $\Omega_Q^*({\frac{3}{2}^-})$,
$\Xi^*_{QQ}({\frac{3}{2}^-})$ and $\Omega^*_{QQ}({\frac{3}{2}^-})$
in Sect.2;
 in Sect.3, we present the numerical results and discussions; and Sect.4 is reserved for our
conclusions.

\section{QCD sum rules for  the negative parity heavy and doubly heavy baryon states}

The ground state quarks have the spin-parity ${1\over 2}^+$, two
quarks can form a scalar diquark or an axialvector diquark with the
spin-parity $0^+$ or $1^+$ without additional relative $P$-wave, the
diquark then combines  with a third quark to form a positive parity
baryon. If there exists a relative $P$-wave (which can be denoted as
$1^-$) between the diquark and the third quark, we can combine the
relative $P$-wave with  the third quark to form an excited quark
with the $J^P={1\over 2}^-$ or ${3\over 2}^-$ firstly, then combine
the negative-parity quark with the positive parity diquark to form a
negative parity baryon. We can denote the quarks, diquarks and
baryons with their spin-parity (or $J^P$), and illustrate the
process  in the following formula  explicitly,
\begin{eqnarray}
\left[{1\over 2}^+\otimes{1\over 2}^+\right]\otimes{1\over 2}^+&=&\left[{0}^{+} +{1}^{+}\right] \otimes{1\over 2}^+={1\over 2}^{+}+\left[{1\over 2}^{+}+{3\over 2}^+\right]\,,\nonumber\\
\left[{1\over 2}^+\otimes{1\over 2}^+\right]\otimes\left[{1\over 2}^+\otimes{1^-}\right]&=&\left[{0}^{+} +{1}^{+}\right] \otimes\left[{1\over 2}^-+{3\over 2}^-\right]\,,\nonumber\\
{1}^{+} \otimes {1\over 2}^+&=&{1\over 2}^++{3\over 2}^+\,,\nonumber\\
{1}^{+} \otimes {1\over 2}^-&=&{1\over 2}^-+{3\over 2}^-\,,
\end{eqnarray}
the baryons with the $0^+$ and $1^+$ diquarks are usually called the
$\Lambda$-type and $\Sigma$-type baryons respectively
\cite{ReviewHB}. In the heavy quark limit, the heavy and doubly
heavy baryon states can be described by the diquark-quark model
\cite{HH-Kiselev}. In the quantum field theory, we usually construct
the $0^+$ and $1^+$ diquark currents with the anti-symmetric Dirac
matrix $C\gamma_5$ and the symmetric Dirac matrix $C\gamma_\mu$,
respectively.  In this article, we construct the $\Sigma$-type
currents to interpolate the ${1\over 2}^{\pm}$ and ${3\over
2}^{\pm}$ heavy and doubly heavy baryon states.

The ${1\over 2}^+$ and ${3\over 2}^+$ heavy and doubly heavy baryon
states    $\Sigma_Q$, $\Xi'_Q$, $\Omega_Q$, $\Xi_{QQ}$,
$\Omega_{QQ}$, $\Sigma_Q^*$, $\Xi_Q^*$, $\Omega_Q^*$, $\Xi^*_{QQ}$
and  $\Omega^*_{QQ}$
 can be interpolated by the following currents
 $J^{\Sigma_Q}(x)$,
$J^{\Xi'_Q}(x)$, $J^{\Omega_Q}(x)$, $J^{\Xi_{QQ}}(x)$,
$J^{\Omega_{QQ}}(x)$, $J^{\Sigma^*_Q}_\mu(x)$, $J^{\Xi^*_Q}_\mu(x)$,
$J^{\Omega^*_Q}_\mu(x)$, $J^{\Xi^*_{QQ}}_\mu(x)$ and
$J^{\Omega^*_{QQ}}_\mu(x)$, respectively,
\begin{eqnarray}
 J^{\Sigma_Q}(x)&=& \epsilon^{ijk}  u^T_i(x)C\gamma_\mu d_j(x)\gamma^\mu \gamma_5Q_k(x)  \, , \nonumber \\
J^{\Xi'_Q}(x)&=& \epsilon^{ijk}  q^T_i(x)C\gamma_\mu s_j(x)   \gamma^\mu \gamma_5Q_k(x)  \, ,  \nonumber \\
J^{\Omega_Q}(x)&=& \epsilon^{ijk}  s^T_i(x)C\gamma_\mu s_j(x)   \gamma^\mu \gamma_5Q_k(x)  \, ,  \nonumber \\
 J^{\Xi_{QQ}}(x)&=& \epsilon^{ijk} Q^T_i(x)C\gamma_\mu Q_j(x) \gamma^\mu \gamma_5q_k(x)  \, ,  \nonumber \\
 J^{\Omega_{QQ}}(x)&=& \epsilon^{ijk}  Q^T_i(x)C\gamma_\mu Q_j(x)   \gamma^\mu \gamma_5s_k(x) \,, \nonumber \\
J^{\Sigma^*_Q}_\mu(x)&=& \epsilon^{ijk}  u^T_i(x)C\gamma_\mu d_j(x)  Q_k(x) \, ,\nonumber \\
J^{\Xi^*_Q}_\mu(x)&=& \epsilon^{ijk}  q^T_i(x)C\gamma_\mu s_j(x)   Q_k(x)  \, ,  \nonumber \\
 J^{\Omega^*_Q}_\mu(x)&=& \epsilon^{ijk}  s^T_i(x)C\gamma_\mu s_j(x)   Q_k(x)  \, ,  \nonumber \\
J^{\Xi^*_{QQ}}_\mu(x)&=& \epsilon^{ijk}  Q^T_i(x)C\gamma_\mu Q_j(x)  q_k(x)  \, ,  \nonumber \\
 J^{\Omega^*_{QQ}}_\mu(x)&=& \epsilon^{ijk} Q^T_i(x)C\gamma_\mu Q_j(x)     s_k(x) \, ,
\end{eqnarray}
where the  $Q$ represents the heavy quarks $c$ and $b$,  the $i$,
$j$ and $k$ are color indexes, and the $C$ is the charge conjunction
matrix.  In this article, we take the Ioffe-type interpolating
currents for the spin $\frac{1}{2}$ baryon states for simplicity
\cite{Ioffe1981}, the general expression of the spin $\frac{1}{2}$
baryon currents is
\begin{eqnarray}
\eta(x)&=&\epsilon^{ijk}\left[\psi_i(x)C\gamma_5\psi_j(x)\psi_k(x)+t\psi_i(x)C\psi_j(x)\gamma_5\psi_k(x)\right]\,
, \nonumber
\end{eqnarray}
where the $\psi_i(x)$ denote the quarks $q$, $s$ and $Q$
\cite{NarisonBook,Bagan93,Espriu1983}. If one choose the optimal
parameter $t$, the results can be improved.

 The corresponding ${1\over 2}^-$ and ${3\over 2}^-$ heavy and doubly heavy baryon states can be
interpolated by the  currents $J^{-} =i\gamma_{5} J^{+}$ and
$J^{-}_\mu =i\gamma_{5} J^{+}_{\mu}$ respectively, because
multiplying $i \gamma_{5}$ to the $J^{+}$ and $J^{+}_\mu$ changes
the parity of the $J^{+}$ and $J^{+}_\mu$ \cite{Oka96}, where the
$J^{+}$ and $J^{+}_\mu$ denote the currents interpolating the
$\Sigma_Q$, $\Xi'_Q$, $\Omega_Q$, $\Xi_{QQ}$, $\Omega_{QQ}$ and
$\Sigma_Q^*$, $\Xi_Q^*$, $\Omega_Q^*$, $\Xi^*_{QQ}$,
$\Omega^*_{QQ}$, respectively.

The correlation functions $\Pi^{\pm}(p)$ and $\Pi^{\pm}_{\mu\nu}(p)$
are defined by
\begin{eqnarray}
\Pi^{\pm}(p)&=&i\int d^4x e^{ip \cdot x} \langle
0|T\left\{J^{\pm}(x)\bar{J}^{\pm}(0)\right\}|0\rangle \, ,\nonumber\\
 \Pi^{\pm}_{\mu\nu}(p)&=&i\int d^4x e^{ip \cdot x}
\langle
0|T\left\{J^{\pm}_\mu(x)\bar{J}^{\pm}_{\nu}(0)\right\}|0\rangle \, .
\end{eqnarray}
The currents $J_{\pm}$ couple  to ${\frac{1}{2}}^{\pm}$ baryon
states $B_{\pm}$, while the currents $J_{\mu}^{\pm}(x)$ couple  to
both the ${\frac{3}{2}}^{\pm}$ baryon states $B^*_{\pm}$ and the
${\frac{1}{2}}^{\pm}$ baryon states $B_{\pm}$ \cite{Chung82}, i.e.
\begin{eqnarray}
    \langle{0}|J_{+}(0)| B_{\pm}(p)\rangle \langle B_{\pm}(p)|\bar{J}_{+}(0)|0\rangle
&=&    - \gamma_{5}\langle 0|J_{-}(0)| B_{\pm}(p)\rangle \langle B_{\pm}(p)| \bar{J}_{-}(0)|0\rangle \gamma_{5} \, , \nonumber \\
\langle{0}|J^{+}_{\mu}(0)| B_{\pm}^*(p)\rangle \langle
B_{\pm}^*(p)|\bar{J}^{+}_{\nu}(0)|0\rangle &=&
    - \gamma_{5}\langle 0|J^{-}_{\mu}(0)| B_{\pm}^*(p)\rangle \langle B_{\pm}^*(p)| \bar{J}^{-}_{\nu}(0)|0\rangle \gamma_{5} \,
, \nonumber \\
    \langle{0}|J^{+}_{\mu}(0)| B_{\pm}(p)\rangle \langle B_{\pm}(p)|\bar{J}^{+}_{\nu}(0)|0\rangle
&=&    - \gamma_{5}\langle 0|J^{-}_{\mu}(0)| B_{\pm}(p)\rangle
\langle B_{\pm}(p)| \bar{J}^{-}_{\nu}(0)|0\rangle \gamma_{5} \, ,
\end{eqnarray}
where
\begin{eqnarray}
\langle 0| J^{\pm} (0)|B_{\pm}(p)\rangle &=&\lambda_{\pm} U(p,s) \, , \nonumber \\
\langle 0| J^{\pm}_\mu (0)|B_{\pm}^*(p)\rangle &=&\lambda_{\pm} U_\mu(p,s) \, , \nonumber \\
 \langle0|J^{\pm}_{\mu}(0)|B_{\mp}(p)\rangle&=&\lambda_{\mp}
 \left(\gamma_{\mu}-4\frac{p_{\mu}}{M_{\mp}}\right)U(p,s) \, ,
\end{eqnarray}
the $\lambda_{\pm}$  are  the  pole residues  and the $M_{\pm}$ are
the masses, and  the spinor $U(p,s)$  satisfies the usual Dirac
equation $(\not\!\!p-M_{\pm})U(p)=0$.

The $\Pi^{\pm}(p)$ and $\Pi^{\pm}_{\mu\nu}(p)$ have the following
relation
\begin{eqnarray}
   \Pi^{-} (p) &=& -\gamma_{5} \Pi^{+} (p)\gamma_{5}   \, , \nonumber \\
\Pi^{-}_{\mu\nu}(p) &=& -\gamma_{5}
\Pi^{+}_{\mu\nu}(p)\gamma_{5} \, .
\end{eqnarray}

We  insert  a complete set  of intermediate baryon states with the
same quantum numbers as the current operators $J^{\pm}(x)$ and
$J_\mu^{\pm}(x)$ into the correlation functions $\Pi^{-}(p)$ and
$\Pi^{-}_{\mu\nu}(p)$ to obtain the hadronic representation
\cite{SVZ79,PRT85}. After isolating the pole terms of the lowest
states of the heavy and doubly heavy baryons, we obtain the
following results  \cite{Oka96}:
\begin{eqnarray}
  \Pi^{-}(p)     & = &  \lambda_{-}^2
    {\!\not\!{p} - M_{-} \over M_{-}^{2}-p^{2}  } +  \lambda_+^2 {\!\not\!{p} +
    M_{+} \over M^{2}_+ -p^{2} }  +\cdots \,
, \nonumber \\
    \Pi^{-}_{\mu\nu}(p)     & = & - \lambda_{-}^2
    {\!\not\!{p} - M_{-} \over M_{-}^{2}-p^{2}  } g_{\mu\nu}-  \lambda_+^2 {\!\not\!{p} +
    M_{+} \over M^{2}_+ -p^{2} }g_{\mu\nu}  +\cdots \,
, \nonumber \\
&=&-\Pi_{-}(p)g_{\mu\nu}+\cdots \, ,
    \end{eqnarray}
where the $M_{\pm}$ are the masses of the lowest states with the
 parity $\pm$ respectively, and the $\lambda_{\pm}$ are the
corresponding pole residues (or couplings). In this article, we
choose the tensor structure $g_{\mu\nu}$ for analysis, the ${1\over
2}^\pm$ baryon states have no contaminations.

 If we take $\vec{p} = 0$ (here we use the $\Pi(p)$ to denote the $\Pi^{-}(p)$ and $\Pi_{-}(p)$ together), then
\begin{eqnarray}
  \rm{limit}_{\epsilon\rightarrow0}\frac{{\rm Im}  \Pi(p_{0}+i\epsilon)}{\pi} & = &
    \lambda_{-}^{2} {\gamma_{0} - 1\over 2} \delta(p_{0} - M_{-})+
    \lambda_+^2 {\gamma_{0} + 1\over 2} \delta(p_{0} - M_+) +\cdots \nonumber \\
  & = & \gamma_{0} A(p_{0}) - B(p_{0})+\cdots \, ,
\end{eqnarray}
where
\begin{eqnarray}
  A(p_{0}) & = &{1 \over 2}\left[\lambda_-^{2} \delta(p_{0} -
  M_{-})+   \lambda_+^{2}
  \delta(p_{0} - M_+)  \right]  \, , \nonumber \\
   B(p_{0}) & = & {1 \over 2} \left[ \lambda_-^{2}
  \delta(p_{0} - M_-)  - \lambda_+^{2} \delta(p_{0} -
  M_{+})\right] \, ,
\end{eqnarray}
the  $A(p_{0}) + B(p_{0})$ and $A(p_{0}) - B(p_{0})$ contain the
contributions  from the negative-parity  baryon states and
positive-parity baryon states,  respectively.

We  calculate the light quark parts of the correlation functions
$\Pi^{-}(p)$ and $\Pi^{-}_{\mu\nu}(p)$ in the coordinate space and
use the momentum space expression for the heavy quark propagators,
i.e. we take
\begin{eqnarray}
S_{ij}(x)&=& \frac{i\delta_{ij}\!\not\!{x}}{ 2\pi^2x^4}
-\frac{\delta_{ij}m_s}{4\pi^2x^2}-\frac{\delta_{ij}}{12}\langle
\bar{s}s\rangle +\frac{i\delta_{ij}}{48}m_s
\langle\bar{s}s\rangle\!\not\!{x}     \nonumber\\
&& -\frac{i}{32\pi^2x^2}  G^{ij}_{\mu\nu}(x) \left[\!\not\!{x}
\sigma^{\mu\nu}+\sigma^{\mu\nu} \!\not\!{x}\right]  +\cdots \, ,\nonumber\\
S_Q^{ij}(x)&=&\frac{i}{(2\pi)^4}\int d^4k e^{-ik \cdot x} \left\{
\frac{\delta_{ij}}{\!\not\!{k}-m_Q}
-\frac{g_sG^{\alpha\beta}_{ij}}{4}\frac{\sigma_{\alpha\beta}(\!\not\!{k}+m_Q)+(\!\not\!{k}+m_Q)
\sigma_{\alpha\beta}}{(k^2-m_Q^2)^2}\right.\nonumber\\
&&\left.+\frac{\pi^2}{3} \langle \frac{\alpha_sGG}{\pi}\rangle
\delta_{ij}m_Q \frac{k^2+m_Q\!\not\!{k}}{(k^2-m_Q^2)^4}
+\cdots\right\} \, ,
\end{eqnarray}
where $\langle \frac{\alpha_sGG}{\pi}\rangle=\langle
\frac{\alpha_sG_{\alpha\beta}G^{\alpha\beta}}{\pi}\rangle$, then
resort to the Fourier integral to transform  the light quark parts
into the momentum space in $D$ dimensions,  take $\vec{p} = 0$,  and
use the dispersion relation to obtain the spectral densities
$\rho^A(p_0)$ and $\rho^B(p_0)$ (which correspond to the tensor
structures $\gamma_0$ and $1$ respectively) at the level of
quark-gluon degrees of freedom. Finally we introduce the weight
functions $\exp\left[-\frac{p_0^2}{T^2}\right]$,
$p_0^2\exp\left[-\frac{p_0^2}{T^2}\right]$,   and obtain the
following sum rules,
\begin{eqnarray}
  \lambda_{-}^2\exp\left[-\frac{M_-^2}{T^2}\right]&=&\int_{\Delta}^{\sqrt{s_0}}dp_0
\left[\rho^A(p_0)
+\rho^B(p_0)\right]\exp\left[-\frac{p_0^2}{T^2}\right] \, ,
\end{eqnarray}
\begin{eqnarray}
  \lambda_{-}^2M_-^2\exp\left[-\frac{M_-^2}{T^2}\right]&=&\int_{\Delta}^{\sqrt{s_0}}dp_0
\left[\rho^A(p_0)
+\rho^B(p_0)\right]p_0^2\exp\left[-\frac{p_0^2}{T^2}\right] \, ,
\end{eqnarray}
where  the $s_0$ are the threshold parameters, the $T^2$ are the
Borel parameters, and $\Delta=2m_Q+m_s$,  $2m_Q$, $m_Q+2m_s$,
$m_Q+m_s$ and $m_Q$ in the channels $\Omega^{(*)}_{QQ}$,
$\Xi^{(*)}_{QQ}$, $\Omega^{(*)}_Q$, $\Xi^{(*)}_Q$ and
$\Sigma_Q^{(*)}$, respectively. The spectral densities $\rho^A(p_0)$
and $\rho^B(p_0)$ at the level of quark-gluon degrees of freedom are
given explicitly in the Appendix.

\section{Numerical results and discussions}
The input parameters are taken to be the standard values $\langle
\bar{q}q \rangle=-(0.24\pm 0.01 \,\rm{GeV})^3$,  $\langle \bar{s}s
\rangle=(0.8\pm 0.2 )\langle \bar{q}q \rangle$, $\langle
\bar{q}g_s\sigma Gq \rangle=m_0^2\langle \bar{q}q \rangle$, $\langle
\bar{s}g_s\sigma Gs \rangle=m_0^2\langle \bar{s}s \rangle$,
$m_0^2=(0.8 \pm 0.2)\,\rm{GeV}^2$ \cite{Ioffe2005,LCSRreview},
$\langle \frac{\alpha_s GG}{\pi}\rangle=(0.012 \pm
0.004)\,\rm{GeV}^4 $ \cite{LCSRreview},
$m_s=(0.14\pm0.01)\,\rm{GeV}$, $m_c=(1.35\pm0.10)\,\rm{GeV}$ and
$m_b=(4.7\pm0.1)\,\rm{GeV}$ \cite{PDG} at the energy scale
$\mu^2=1\, \rm{GeV}^2$.

The value of the gluon condensate $\langle \frac{\alpha_s
GG}{\pi}\rangle $ has been updated from time to time, and changes
greatly \cite{NarisonBook}.
 At the present case, the gluon condensate  makes tiny  contribution,
the updated value $\langle \frac{\alpha_s GG}{\pi}\rangle=(0.023 \pm
0.003)\,\rm{GeV}^4 $ \cite{NarisonBook} and the standard value
$\langle \frac{\alpha_s GG}{\pi}\rangle=(0.012 \pm
0.004)\,\rm{GeV}^4 $ \cite{LCSRreview} lead to a tiny  difference
and can be neglected safely.

For the light quark masses, we
 take the approximation $m_u=m_d\approx0$. The values listed in the Review of Particle
 Physics are  $m_u=2.49^{+0.81}_{-0.79}\,\rm{MeV}$ and
 $m_d=5.05^{+0.75}_{-0.95}\,\rm{MeV}$ at the energy scale $\mu=2\,\rm{GeV}$
 \cite{PDG}. The values  $m_q=\frac{m_u+m_d}{2}=5.6\,\rm{MeV}$
 \cite{NarisonBook} and $m_q=0$
lead to a difference  less
 than $2\,\rm{MeV}$ for masses of the  $Qqs$, $Qqq^{\prime}$
and $QQq$ baryon states, we can neglect the $u$ and $d$ quark masses
safely.

The $Q$-quark masses appearing in the perturbative terms  are
usually taken to be the pole masses in the QCD sum rules, while the
choice of the $m_Q$ in the leading-order coefficients of the
higher-dimensional terms is arbitrary \cite{NarisonBook,Kho9801}.
The $\overline{MS}$ mass $m_c(m_c^2)$ relates with the pole mass
$\hat{m}_c$ through the relation $ m_c(m_c^2)
=\hat{m}_c\left[1+\frac{C_F \alpha_s(m_c^2)}{\pi}+\cdots\right]^{-1}
$. In this article, we take the approximation
$m_c(m_c^2)\approx\hat{m}_c$ without the $\alpha_s$ corrections for
consistency. The value listed in the Review of Particle Physics is
$m_c(m_c^2)=1.27^{+0.07}_{-0.09} \, \rm{GeV}$ \cite{PDG}, it is
reasonable to take
$\hat{m}_c=m_c(1\,\rm{GeV}^2)=(1.35\pm0.10)\,\rm{GeV}$. For the $b$
quark,  the $\overline{MS}$ mass is
$m_b(m_b^2)=4.19^{+0.18}_{-0.06}\,\rm{GeV}$ \cite{PDG}, the
  gap between the energy scale $\mu=4.2\,\rm{GeV}$ and
 $1\,\rm{GeV}$ is rather large, the approximation $\hat{m}_b\approx m_b(m_b^2)\approx m_b(1\,\rm{GeV}^2)$ seems rather crude.
  It would be better to understand the heavy quark masses $m_c$ and $m_b$ we
take at the energy scale $\mu^2=1\,\rm{GeV}^2$ as the effective
quark masses (or just the mass parameters). Our previous works on
the mass spectrum of the ${\frac{1}{2}}^+$ and ${\frac{3}{2}}^+$
heavy and doubly heavy baryon states and the ${\frac{1}{2}}^-$
antitriplet heavy baryon states indicate such parameters can lead to
satisfactory results \cite{Wang0912,Wang0101,Wang0101-2,Wang0102}.

In Ref.\cite{HH-Narison},  Bagan et al observe that the contribution
from the mixed condensates has a term that behaves like
$\frac{1}{v^3}$ (the $v$ is the heavy quark velocity) for the doubly
heavy baryon states, which indicates a coulombic correction and
 requires a complete treatment of the nonrelativistic coulombic
corrections. In this article, we take (almost) the same pole
contributions and analogous convergent behaviors in the operator
product expansion as our previous works \cite{Wang0101-2,Wang0102},
where  the positive-parity doubly heavy baryon states were studied
with the vacuum condensates up to dimension 4, and neglect the
contributions  from the mixed condensates for consistency.  We also
present the results with the mixed condensates  for comparison, see
Table 1. For
  the ${\frac{1}{2}}^-$ doubly heavy baryon states,
there are  cancellations among the mixed condensates which originate
from the diagrams where  the gluons emitted from
  the two heavy quarks and the light quark respectively and
 then absorbed by the light quark.   The net contributions are minor and can be
neglected safely. On the other hand, for the ${\frac{3}{2}}^-$
doubly heavy  baryon states, there are only contributions from the
mixed condensates which originate from the diagrams where  the
gluons emitted from the light quark  and then absorbed by the light
quark itself. There are no cancellations,  the inclusion of the
mixed condensates can result in masses about $(30-50)\,\rm{MeV}$
larger than the corresponding ones without  them, and the
uncertainties  of the
 masses originate from the parameter $m_0^2=(0.8\pm0.2)\,\rm{GeV}^2$
are less than $10\,\rm{MeV}$.

In calculations, we   neglect  the contributions from the
perturbative corrections $\mathcal {O}(\alpha_s^n)$.  Those
perturbative corrections can be taken into account in the leading
logarithmic
 approximations through  anomalous dimension factors. After the Borel transform, the effects of those
 corrections are  to multiply each term on the operator product
 expansion side by the factor, $ \left[ \frac{\alpha_s(T^2)}{\alpha_s(\mu^2)}\right]^{2\Gamma_{J}-\Gamma_{\mathcal
 {O}_n}}  $,
 where the $\Gamma_{J}$ is the anomalous dimension of the
 interpolating current $J(x)$ and the $\Gamma_{\mathcal {O}_n}$ is the anomalous dimension of
 the local operator $\mathcal {O}_n(0)$. We carry out the operator product expansion at a special energy
scale $\mu^2=1\,\rm{GeV}^2$, and  set the factor $\left[
\frac{\alpha_s(T^2)}{\alpha_s(\mu^2)}\right]^{2\Gamma_{J}-\Gamma_{\mathcal
{O}_n}}\approx1$, such an approximation maybe result in some scale
dependence  and  weaken the prediction ability. In this article, we
study the $\frac{1}{2}^-$ and $\frac{3}{2}^-$ heavy and doubly heavy
baryon states systemically,  the predictions are still robust   as
we take the analogous criteria in those sum rules.

 In Refs.\cite{Wang0912,Wang0101,Wang0101-2,Wang0102}, we study  the
 masses of the ${\frac{1}{2}}^{\pm}$ antitriplet heavy baryon states $\Lambda_Q({\frac{1}{2}}^{\pm})$, $\Xi_Q({\frac{1}{2}}^{\pm})$, and the
${\frac{1}{2}}^+$ and ${\frac{3}{2}}^+$ heavy and doubly heavy
baryon states $\Sigma_Q$, $\Xi_Q^{\prime}$, $\Omega_Q$, $\Xi_{QQ}$,
$\Omega_{QQ}$, $\Sigma_Q^*$, $\Xi_Q^*$, $\Omega_Q^*$, $\Xi^*_{QQ}$
and $\Omega^*_{QQ}$
 systematically  by subtracting the contributions of the
corresponding  negative (or positive) parity heavy and doubly heavy
baryon states, and obtain satisfactory results. In calculations, we
take the values $\sqrt{s_0}-M_{\rm{gr}}\approx (0.7-0.8)\,\rm{GeV}$
and the pole contributions $\approx$ $(45-80)\%$ and $(45-70)\%$ in
the charmed channels and the bottom channels respectively, where the
$s_0$ stands for the central values of the threshold parameters, and
the $M_{\rm gr}$ stands for the experimental values of the heavy and
doubly heavy baryon states $\Lambda_{c}({\frac{1}{2}}^{\pm})$,
$\Xi_{c}({\frac{1}{2}}^{\pm})$, $\Lambda_{b}$, $\Xi_{b}$,
$\Sigma_c$, $\Xi_c^{\prime}$, $\Omega_c$, $\Sigma_b$,  $\Omega_b$,
$\Sigma_c^*$, $\Xi_c^{*}$, $\Omega_c^*$, $\Sigma_b^*$, $\Omega_{cc}$
listed in the Review of Particle Physics \cite{PDG}, here we take
the average value of the mass of the $\Omega_b$ from the
experimental data of the D0 and CDF collaborations
\cite{OmegabD0,OmegabCDF}.

In the conventional QCD sum rules \cite{SVZ79,PRT85}, there are two
criteria (pole dominance and convergence of the operator product
expansion) for choosing  the Borel parameters $T^2$ and threshold
parameters $s_0$.  We can borrow some ideas from our previous works
\cite{Wang0912,Wang0101,Wang0101-2,Wang0102}, and take the values
$\sqrt{s_0}=M_{\rm gr}+0.7\,\rm{GeV}$ as guides (where the $M_{\rm
gr}$ denotes the ground state masses of the negative parity heavy
and doubly heavy baryon states from  the  non-relativistic quark
model \cite{HH-Roberts}),  and impose the two criteria on the heavy
and doubly heavy baryon states to choose the Borel parameters $T^2$
and threshold parameters $s_0$. If we take the contributions of the
pole terms to be  $(45-80)\%$ and $(45-70)\%$ in the charmed
channels and the bottom channels respectively (for the central
values of the threshold parameters, the pole contributions are
larger than or equal $50\%$), and the contributions of the
perturbative terms to be $(50-90\%)$ in the operator product
expansion as in our previous works
\cite{Wang0912,Wang0101,Wang0101-2,Wang0102}, we can obtain
approximately uniform Borel windows. In this article, we take the
uniform Borel windows $T^2_{max}-T^2_{min}=1.0 \, \rm{GeV}^2$, $1.5
\, \rm{GeV}^2$ and $2.0 \, \rm{GeV}^2$ in the singly heavy baryon
channels, the doubly charmed baryon channels  and the doubly bottom
baryon channels, respectively; the corresponding threshold
parameters are presented in Table 1. The two criteria of the QCD sum
rules are fully satisfied \cite{SVZ79,PRT85}. If we take the values
$T^2<T^2_{min}$, the resulting  masses and pole residues change
quickly with variations of the Borel parameters, and no satisfactory
Borel plateaus can be obtained. On the other hand, if we take the
values  $T^2>T^2_{max}$, the pole contributions are reduced
remarkably, we have to postpone the threshold parameters $s_0$ to
larger values, the contributions from the higher resonances and
continuum states may be  included in, which will weaken the
prediction ability. Although with fine-tuning, we can obtain
slightly larger Borel windows, the predictions cannot be improved
remarkably, we prefer the uniform Borel windows $1\,\rm{GeV}^2$,
$1.5\,\rm{GeV}^2$ and $2.0\,\rm{GeV}^2$ in the due channels.

In this article, we take uniform uncertainties for the threshold
parameters, $\delta_{\sqrt{s_0}}=\pm 0.1\, \rm{GeV}$. In
calculations, we observe that the   predicted masses are not
sensitive to the threshold parameters in the Borel windows, although
they increase with the threshold parameters, the higher resonances
and continuum states are suppressed by the factor
$\exp\left[-\frac{p_0^2}{T^2}\right]\Theta(p_0^2-s_0)\leq e^{-1}$.
In fact, we can take larger uncertainties,  $\delta_{\sqrt{s_0}}=\pm
0.2\, \rm{GeV}$, which can enlarge the uncertainties of the masses
about $(0.01-0.03)\,\rm{GeV}$. For example, if we take the larger
uncertainties $\delta_{\sqrt{s_0}}=\pm 0.2\, \rm{GeV}$  and the same
pole contributions, the Borel parameters in the channels $\Omega_c$,
$\Omega^*_c$, $\Xi^{\prime}_c$ and $\Xi_c^*$ should be taken as
$T^2\approx(2.9-3.5)\,\rm{GeV}^2$, $(2.9-3.5)\,\rm{GeV}^2$,
$(2.7-3.3)\,\rm{GeV}^2$ and $(2.7-3.3)\,\rm{GeV}^2$, respectively,
i.e. the Borel windows shrink,
  the predicted masses are
 $M=(2.98\pm0.18)\,\rm{GeV}$, $(2.98\pm0.18)\,\rm{GeV}$, $(2.87\pm0.18)\,\rm{GeV}$ and
 $(2.87\pm0.19)\,\rm{GeV}$, respectively; the uncertainties are
 enlarged by  $(0.01-0.03)\,\rm{GeV}$.

Taking into account all uncertainties  of the revelent  parameters,
we can obtain the values of the masses and pole residues of
 the ${1\over 2}^-$ and ${3\over 2}^-$ heavy and doubly heavy
baryon states $\Sigma_Q({\frac{1}{2}^-})$,
$\Xi'_Q({\frac{1}{2}^-})$, $\Omega_Q({\frac{1}{2}^-})$,
$\Xi_{QQ}({\frac{1}{2}^-})$, $\Omega_{QQ}({\frac{1}{2}^-})$,
$\Sigma_Q^*({\frac{3}{2}^-})$, $\Xi_Q^*({\frac{3}{2}^-})$,
$\Omega_Q^*({\frac{3}{2}^-})$, $\Xi^*_{QQ}({\frac{3}{2}^-})$ and
$\Omega^*_{QQ}({\frac{3}{2}^-})$, which are shown in Figs.1-4 and
Table  1. In this article,  we calculate the uncertainties $\delta$
with the formula
\begin{eqnarray}
\delta=\sqrt{\sum_i\left(\frac{\partial f}{\partial
x_i}\right)^2\mid_{x_i=\bar{x}_i} (x_i-\bar{x}_i)^2}\,  ,
\end{eqnarray}
 where the $f$ denotes  the
hadron mass  $M_{-}$ and the pole residue $\lambda_{-}$,  the $x_i$
denotes  the input QCD parameters $m_c$, $m_b$, $\langle \bar{q}q
\rangle$, $\langle \bar{s}s \rangle$, $\cdots$, and the threshold
parameter $s_0$ and Borel parameter $M^2$. As the partial
 derivatives   $\frac{\partial f}{\partial x_i}$ are difficult to carry
out analytically, we take the  approximation $\left(\frac{\partial
f}{\partial x_i}\right)^2 (x_i-\bar{x}_i)^2\approx
\left[f(\bar{x}_i\pm \Delta x_i)-f(\bar{x}_i)\right]^2$ in the
numerical calculations.

In Table 1, we also present the results from the non-relativistic
quark model \cite{HH-Roberts} and the relativistic quark model based
on a quasipotential approach in QCD \cite{HH-Ebert-1,HH-Ebert-2},
the present  predictions are in good agreement with those values
within uncertainties.

\begin{table}
\begin{center}
\begin{tabular}{|c|c|c|c|c|c|c|c|}\hline\hline
                                              & $T^2 (\rm{GeV}^2)$ & $\sqrt{s_0} (\rm{GeV})$ & $M(\rm{GeV})$   & $\lambda(\rm{GeV}^3)$  & Ref.\cite{HH-Roberts}  & Refs.\cite{HH-Ebert-1,HH-Ebert-2}\\ \hline
                  $\Sigma_c({\frac{1}{2}^-})$ & $2.3-3.3$          & $3.5\pm0.1$             & $2.74\pm0.20$   & $0.071\pm0.019$        & $2.748$                & $2.795$\\ \hline
                    $\Xi'_c({\frac{1}{2}^-})$ & $2.5-3.5$          & $3.6\pm0.1$             & $2.87\pm0.17$   & $0.084\pm0.019$        & $2.859$                & $2.928$\\ \hline
                  $\Omega_c({\frac{1}{2}^-})$ & $2.7-3.7$          & $3.7\pm0.1$             & $2.98\pm0.16$   & $0.136\pm0.027$        & $2.977$                & $3.020$\\ \hline
                  $\Sigma_b({\frac{1}{2}^-})$ & $4.9-5.9$          & $6.8\pm0.1$             & $6.00\pm0.18$   & $0.085\pm0.022$        & $6.099$                & $6.108$\\ \hline
                    $\Xi'_b({\frac{1}{2}^-})$ & $5.2-6.2$          & $6.9\pm0.1$             & $6.14\pm0.15$   & $0.103\pm0.024$        & $6.192$                & $6.238$\\ \hline
                  $\Omega_b({\frac{1}{2}^-})$ & $5.5-6.5$          & $7.0\pm0.1$             & $6.27\pm0.14$   & $0.173\pm0.035$        & $6.301$                & $6.352$\\ \hline
                  $\Xi_{cc}({\frac{1}{2}^-})$ & $3.1-4.6$          & $4.5\pm0.1$             & $3.77\pm0.18$   & $0.159\pm0.037$        & $3.910$                & $3.838$\\ \hline
        $\widehat{\Xi_{cc}}({\frac{1}{2}^-})$ & $3.1-4.6$          & $4.5\pm0.1$             & $3.78\pm0.18$   & $0.160\pm0.037$        & $3.910$                & $3.838$\\ \hline
               $\Omega_{cc}({\frac{1}{2}^-})$ & $3.4-4.9$          & $4.6\pm0.1$             & $3.91\pm0.14$   & $0.192\pm0.041$        & $4.046$                & $4.002$\\ \hline
     $\widehat{\Omega_{cc}}({\frac{1}{2}^-})$ & $3.4-4.9$          & $4.6\pm0.1$             & $3.92\pm0.16$   & $0.192\pm0.041$        & $4.046$                & $4.002$\\ \hline
                  $\Xi_{bb}({\frac{1}{2}^-})$ & $8.8-10.8$         & $11.1\pm0.1$            & $10.38\pm0.15$  & $0.364\pm0.088$        & $10.493$               & $10.632$\\ \hline
        $\widehat{\Xi_{bb}}({\frac{1}{2}^-})$ & $8.8-10.8$         & $11.1\pm0.1$            & $10.39\pm0.15$  & $0.365\pm0.089$        & $10.493$               & $10.632$\\ \hline
               $\Omega_{bb}({\frac{1}{2}^-})$ & $9.1-11.1$         & $11.2\pm0.1$            & $10.53\pm0.15$  & $0.443\pm0.101$        & $10.616$               & $10.771$\\ \hline
     $\widehat{\Omega_{bb}}({\frac{1}{2}^-})$ & $9.1-11.1$         & $11.2\pm0.1$            & $10.53\pm0.15$  & $0.444\pm0.101$        & $10.616$               & $10.771$\\ \hline
                $\Sigma^*_c({\frac{3}{2}^-})$ & $2.4-3.4$          & $3.5\pm0.1$             & $2.74\pm0.20$   & $0.037\pm0.009$        & $2.763$                & $2.761$\\ \hline
                   $\Xi^*_c({\frac{3}{2}^-})$ & $2.6-3.6$          & $3.6\pm0.1$             & $2.86\pm0.17$   & $0.045\pm0.009$        & $2.871$                & $2.900$\\ \hline
                $\Omega^*_c({\frac{3}{2}^-})$ & $2.8-3.8$          & $3.7\pm0.1$             & $2.98\pm0.16$   & $0.072\pm0.013$        & $2.986$                & $2.998$\\ \hline
                $\Sigma^*_b({\frac{3}{2}^-})$ & $5.0-6.0$          & $6.8\pm0.1$             & $6.00\pm0.18$   & $0.047\pm0.012$        & $6.101$                & $6.076$\\ \hline
                   $\Xi^*_b({\frac{3}{2}^-})$ & $5.3-6.3$          & $6.9\pm0.1$             & $6.14\pm0.16$   & $0.054\pm0.013$        & $6.194$                & $6.212$\\ \hline
                $\Omega^*_b({\frac{3}{2}^-})$ & $5.6-6.6$          & $7.0\pm0.1$             & $6.26\pm0.15$   & $0.095\pm0.019$        & $6.304$                & $6.330$\\  \hline
                $\Xi^*_{cc}({\frac{3}{2}^-})$ & $3.3-4.8$          & $4.5\pm0.1$             & $3.77\pm0.17$   & $0.087\pm0.019$        & $3.921$                & $3.959$\\ \hline
      $\widehat{\Xi^*_{cc}}({\frac{3}{2}^-})$ & $3.5-5.0$          & $4.6\pm0.1$             & $3.80\pm0.18$   & $0.095\pm0.020$        & $3.921$                & $3.959$\\ \hline
             $\Omega^*_{cc}({\frac{3}{2}^-})$ & $3.6-5.1$          & $4.6\pm0.1$             & $3.91\pm0.16$   & $0.105\pm0.020$        & $4.052$                & $4.102$\\ \hline
   $\widehat{\Omega^*_{cc}}({\frac{3}{2}^-})$ & $3.9-5.4$          & $4.7\pm0.1$             & $3.96\pm0.16$   & $0.116\pm0.022$        & $4.052$                & $4.102$\\ \hline
                $\Xi^*_{bb}({\frac{3}{2}^-})$ & $9.0-11.0$         & $11.1\pm0.1$            & $10.39\pm0.15$  & $0.206\pm0.049$        & $10.495$               & $10.647$\\ \hline
      $\widehat{\Xi^*_{bb}}({\frac{3}{2}^-})$ & $9.3-11.3$         & $11.2\pm0.1$            & $10.43\pm0.15$  & $0.227\pm0.052$        & $10.495$               & $10.647$\\ \hline
             $\Omega^*_{bb}({\frac{3}{2}^-})$ & $9.3-11.3$         & $11.2\pm0.1$            & $10.52\pm0.15$  & $0.251\pm0.056$        & $10.619$               & $10.785$\\ \hline
   $\widehat{\Omega^*_{bb}}({\frac{3}{2}^-})$ & $9.6-11.6$         & $11.3\pm0.1$            & $10.57\pm0.15$  & $0.275\pm0.059$        & $10.619$               & $10.785$\\ \hline
   \hline
\end{tabular}
\end{center}
\caption{ The masses and pole residues  of   the ${\frac{1}{2}^-}$
and ${\frac{3}{2}^-}$ heavy and doubly heavy baryon states, we also
present the predictions for the masses from some quark models for
comparison. The wide-hat \, $\widehat{}$ \,  denotes that the mixed
condensates are taken into account. }
\end{table}

\begin{figure}
 \centering
 \includegraphics[totalheight=3.5cm,width=3.5cm]{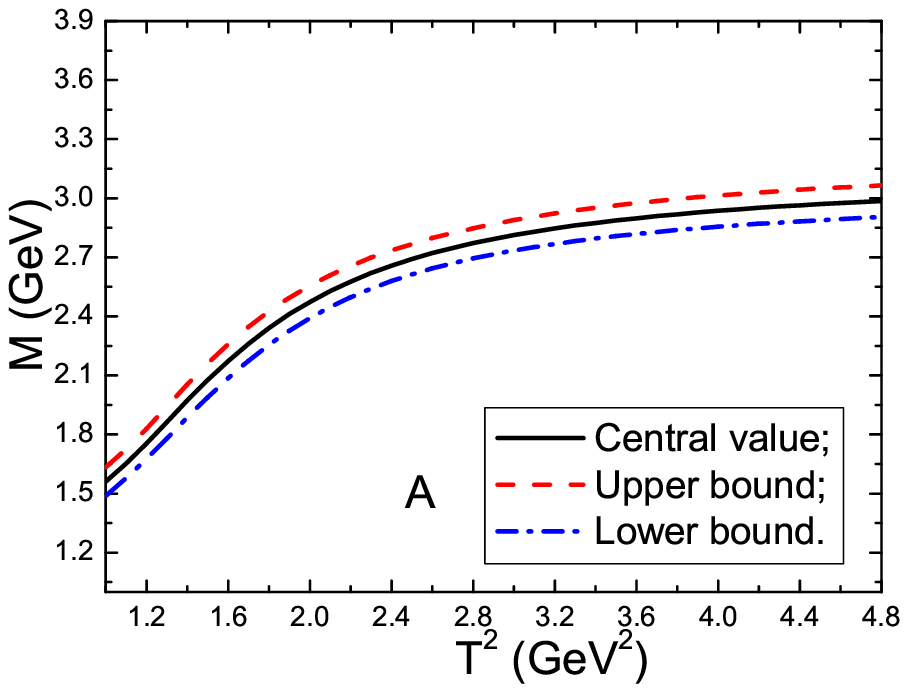}
 \includegraphics[totalheight=3.5cm,width=3.5cm]{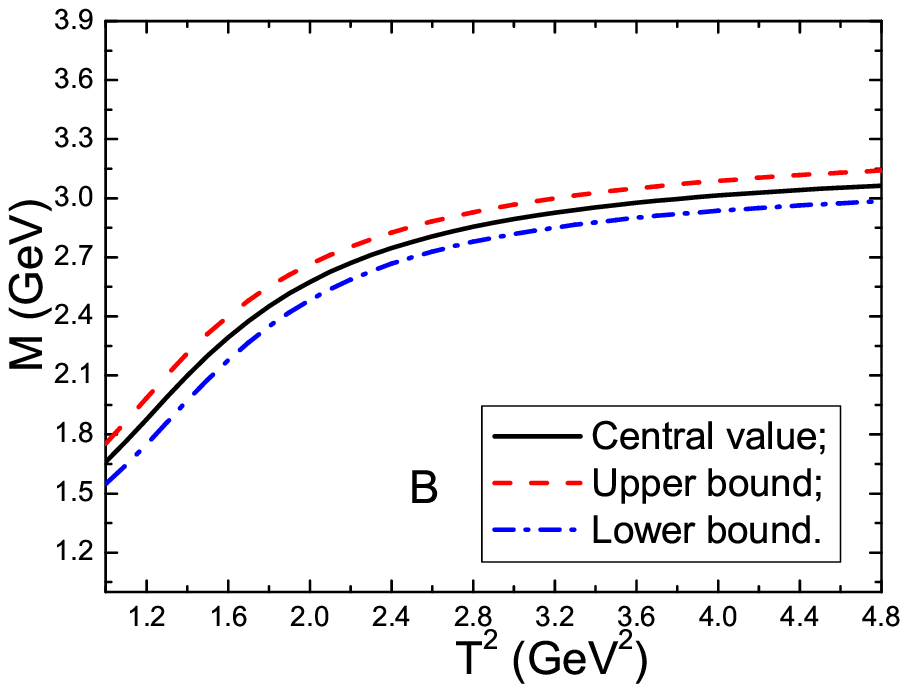}
 \includegraphics[totalheight=3.5cm,width=3.5cm]{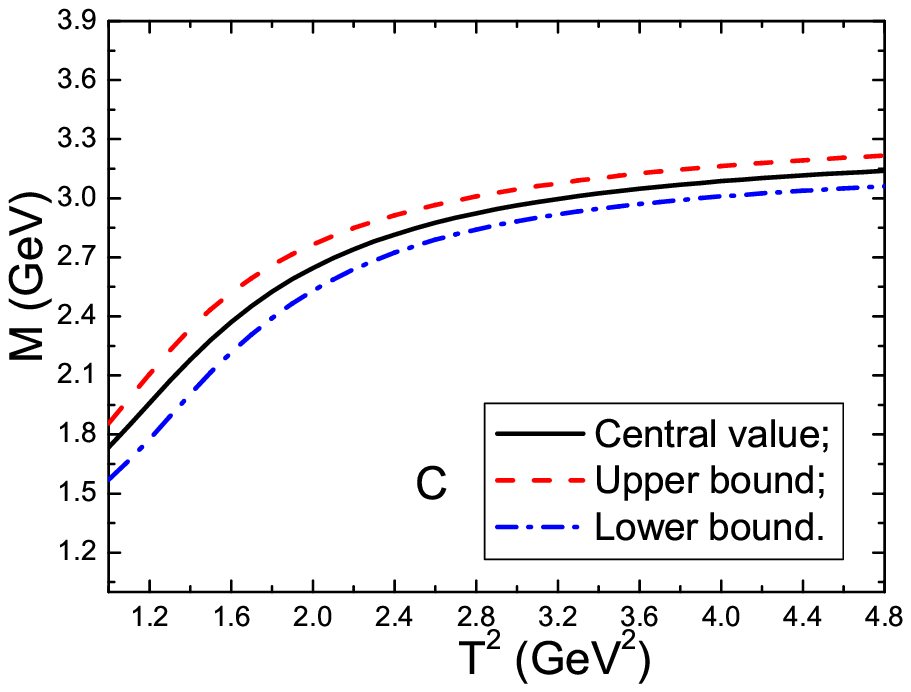}
 \includegraphics[totalheight=3.5cm,width=3.5cm]{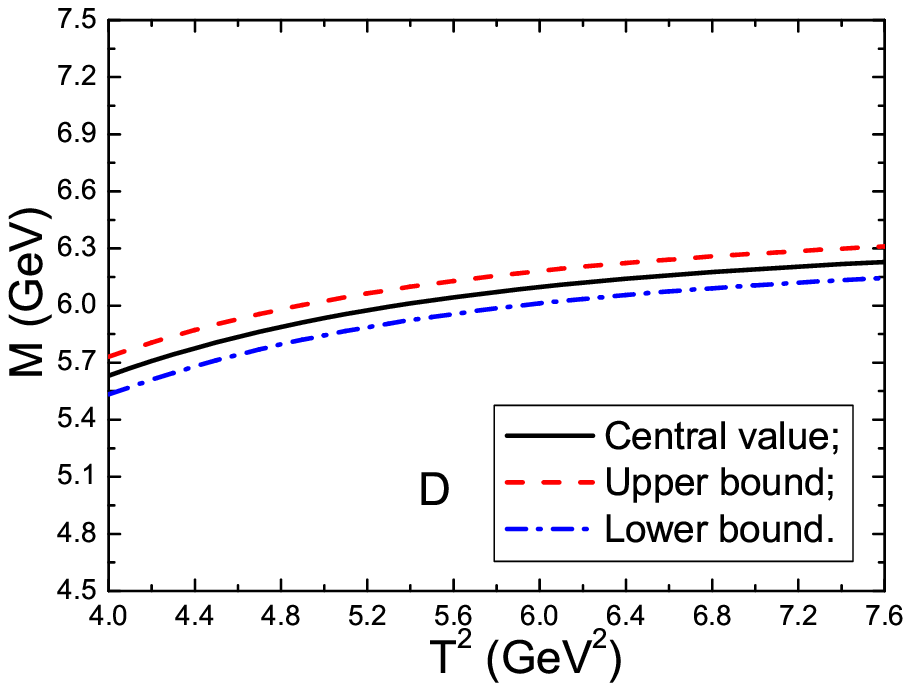}
 \includegraphics[totalheight=3.5cm,width=3.5cm]{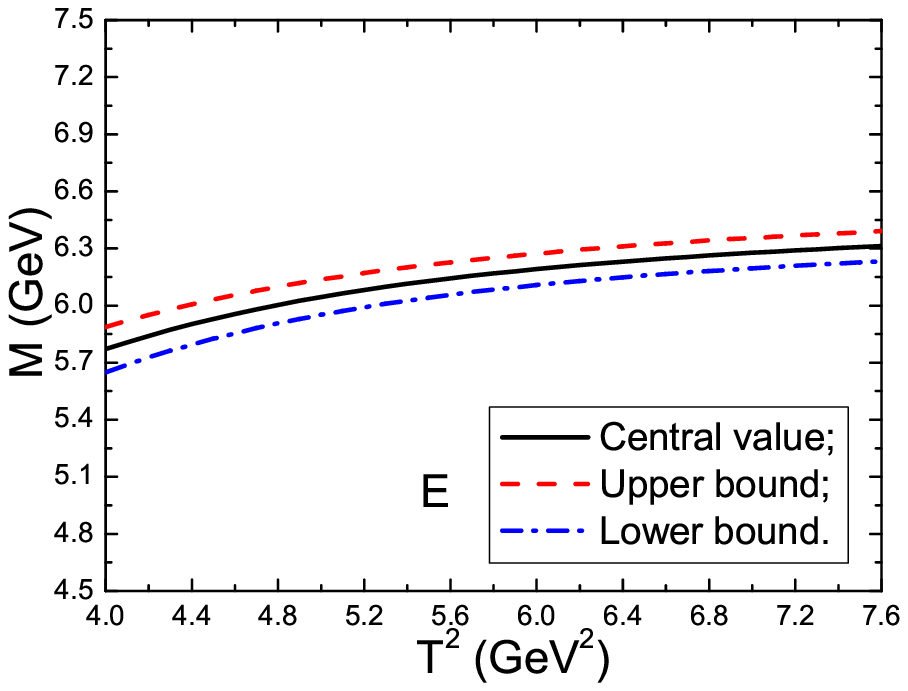}
 \includegraphics[totalheight=3.5cm,width=3.5cm]{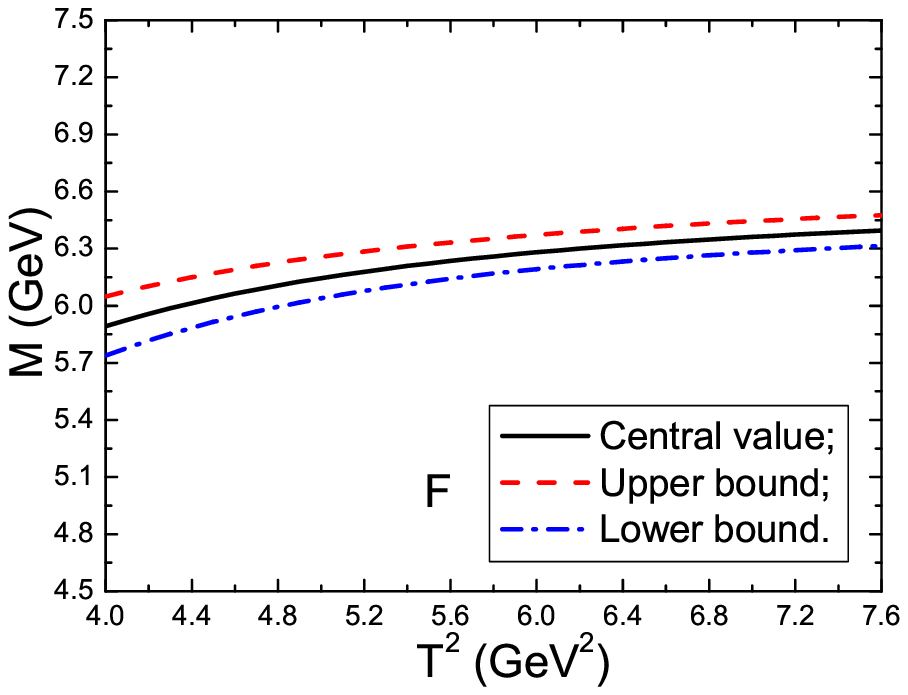}
 \includegraphics[totalheight=3.5cm,width=3.5cm]{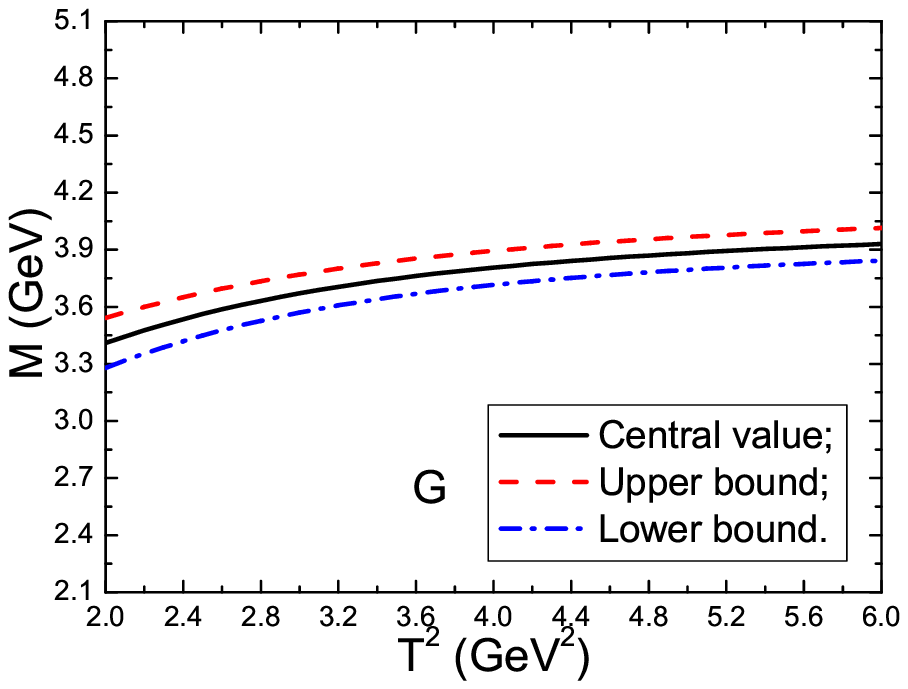}
 \includegraphics[totalheight=3.5cm,width=3.5cm]{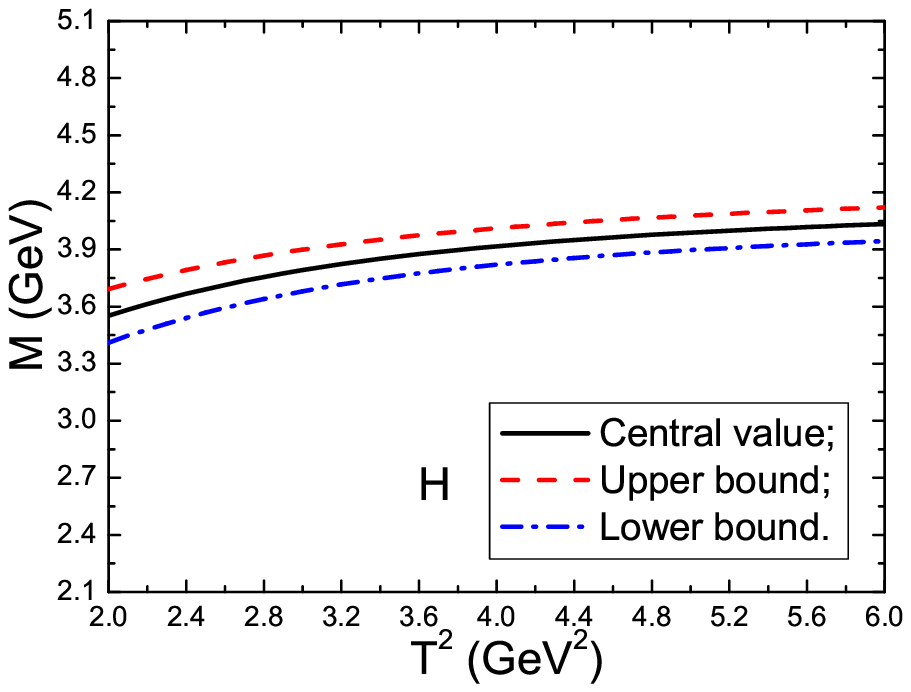}
 \includegraphics[totalheight=3.5cm,width=3.5cm]{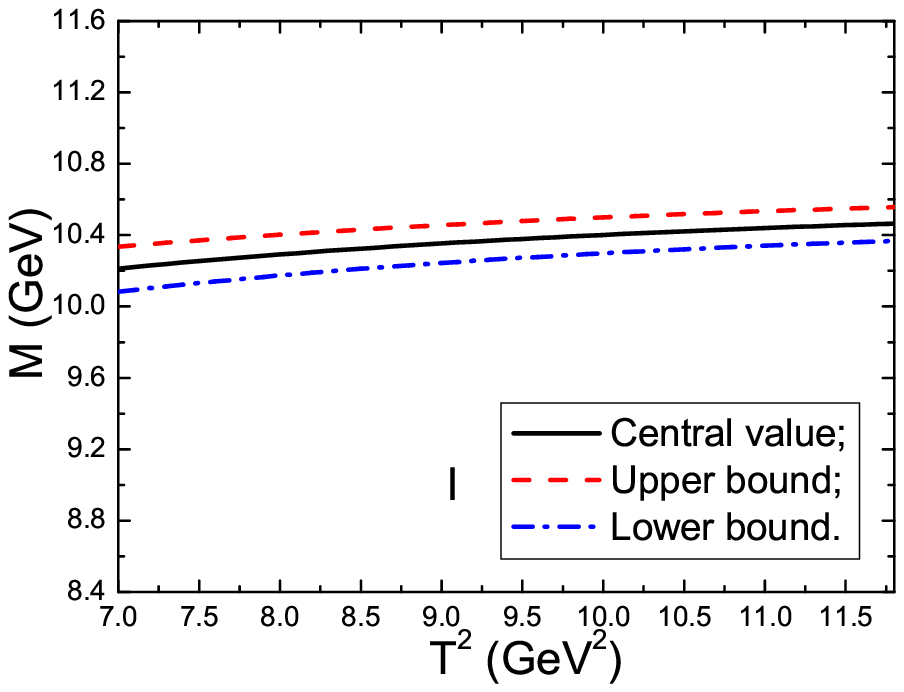}
 \includegraphics[totalheight=3.5cm,width=3.5cm]{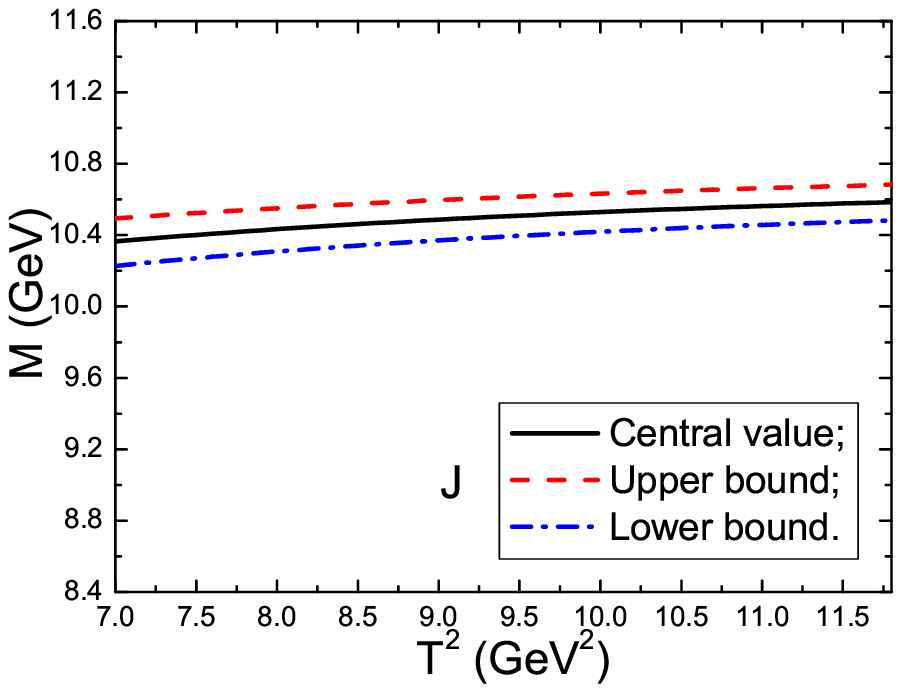}
  \caption{ The  masses of the ${\frac{1}{2}}^-$ heavy and doubly heavy baryon states, the $A$, $B$, $C$, $D$, $E$, $F$, $G$, $H$, $I$ and $J$  correspond
     to the channels $\Sigma_c({1\over 2}^-)$, $\Xi'_c({1\over 2}^-)$, $\Omega_c({1\over 2}^-)$, $\Sigma_b({1\over 2}^-)$, $\Xi'_b({1\over 2}^-)$, $\Omega_b({1\over 2}^-)$,
  $\Xi_{cc}({1\over 2}^-)$, $\Omega_{cc}({1\over 2}^-)$, $\Xi_{bb}({1\over 2}^-)$ and $\Omega_{bb}({1\over 2}^-)$, respectively.  }
\end{figure}

\begin{figure}
 \centering
 \includegraphics[totalheight=3.5cm,width=3.5cm]{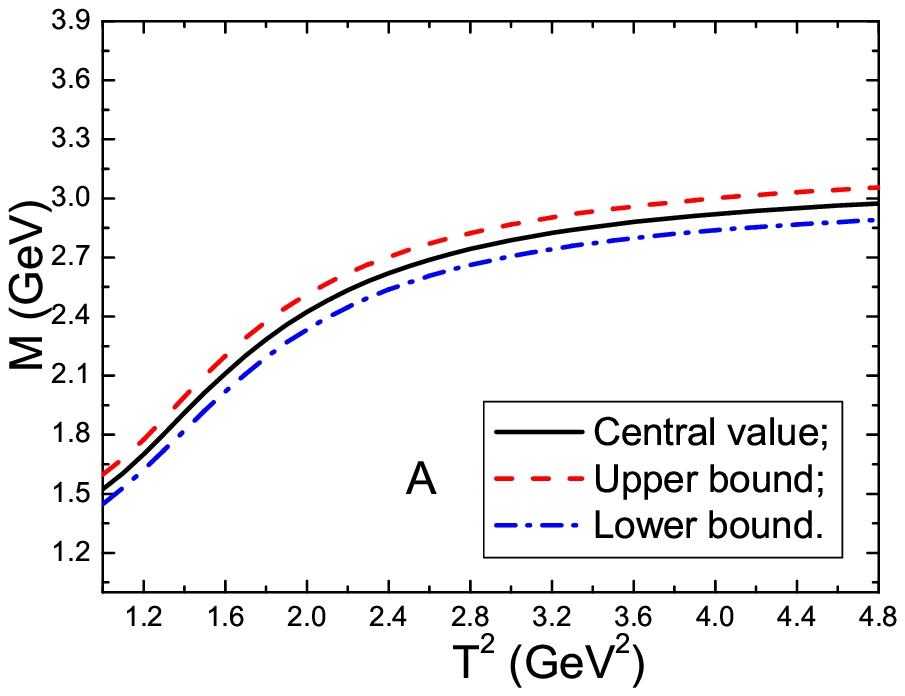}
 \includegraphics[totalheight=3.5cm,width=3.5cm]{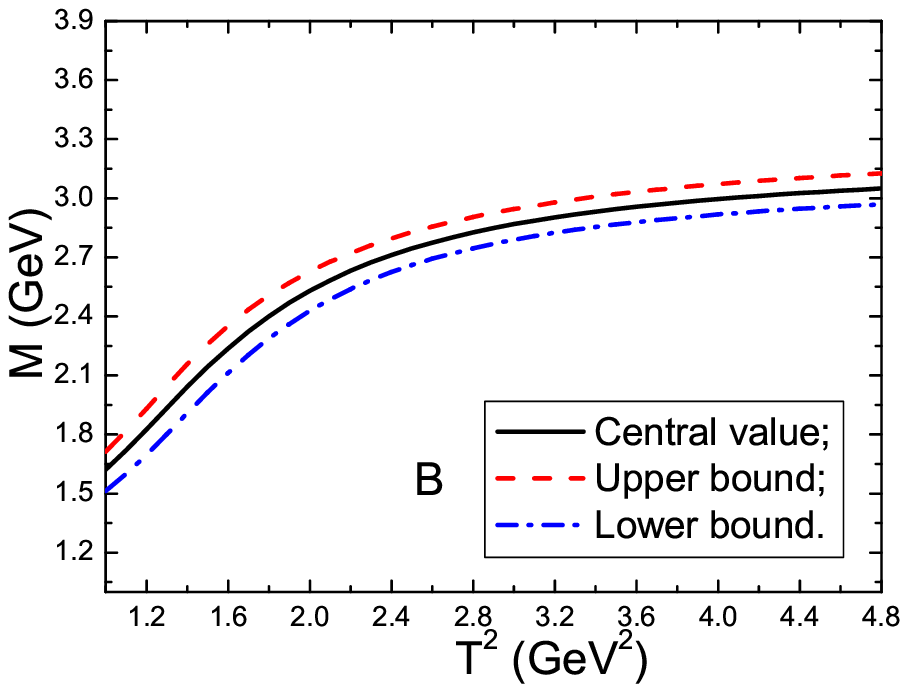}
 \includegraphics[totalheight=3.5cm,width=3.5cm]{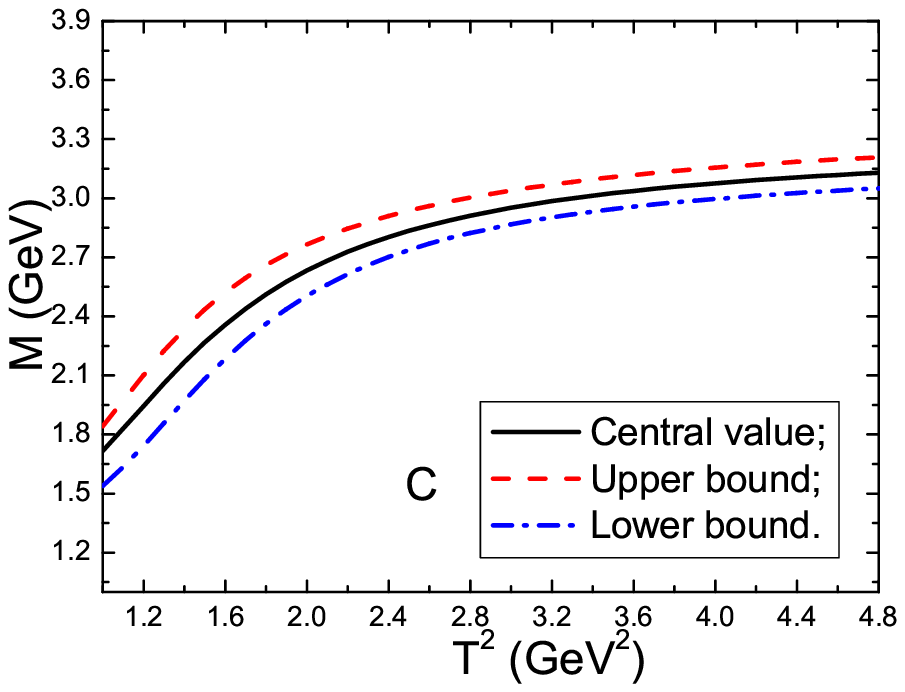}
 \includegraphics[totalheight=3.5cm,width=3.5cm]{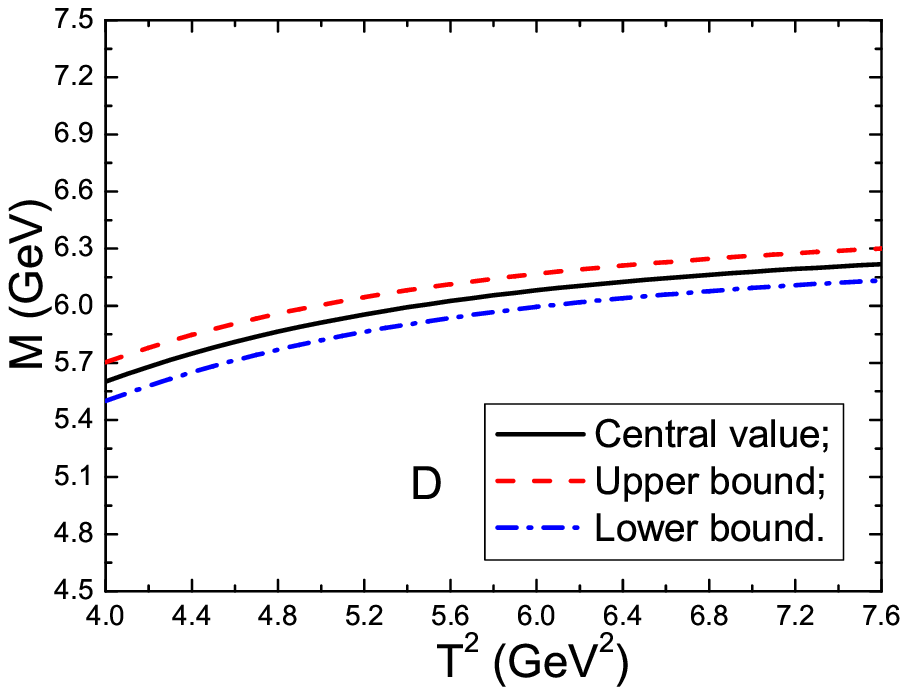}
 \includegraphics[totalheight=3.5cm,width=3.5cm]{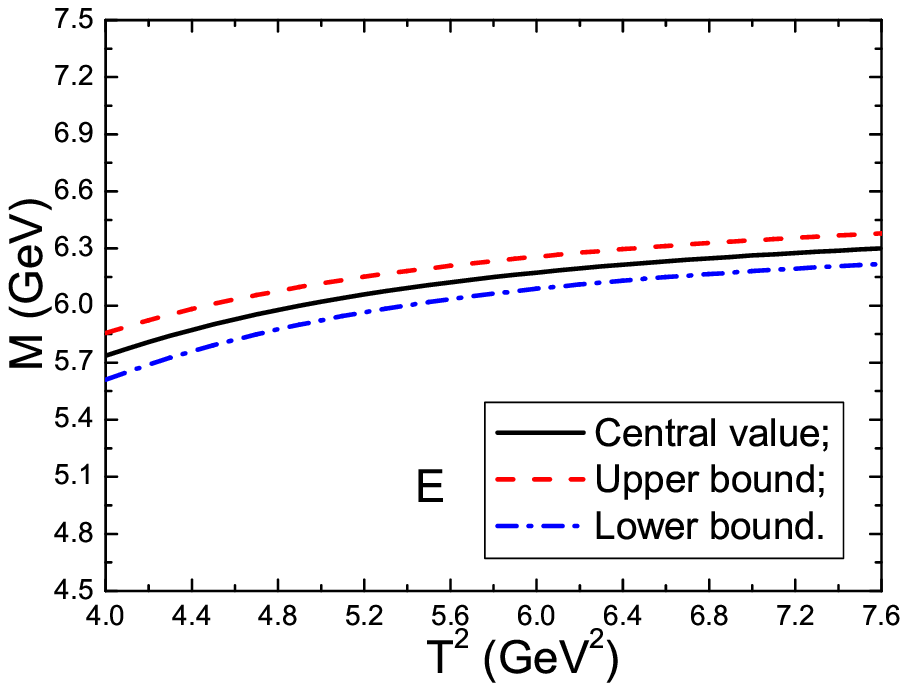}
 \includegraphics[totalheight=3.5cm,width=3.5cm]{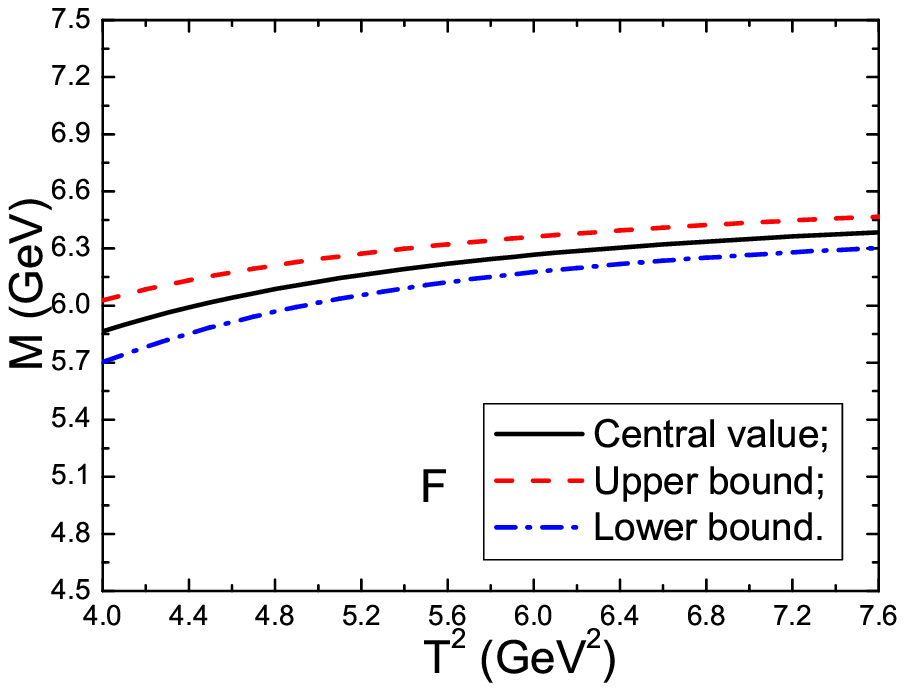}
 \includegraphics[totalheight=3.5cm,width=3.5cm]{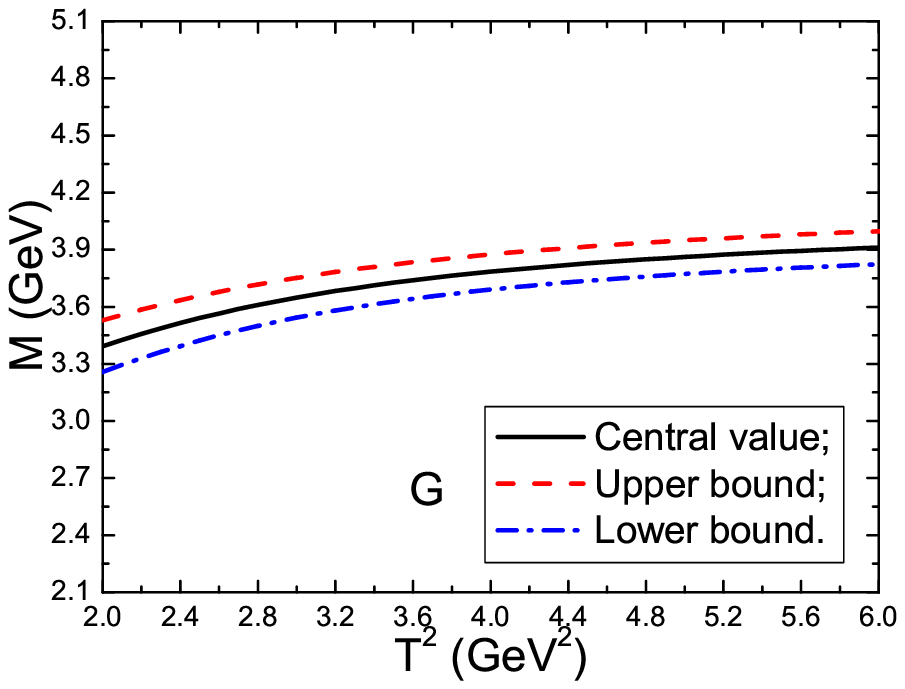}
 \includegraphics[totalheight=3.5cm,width=3.5cm]{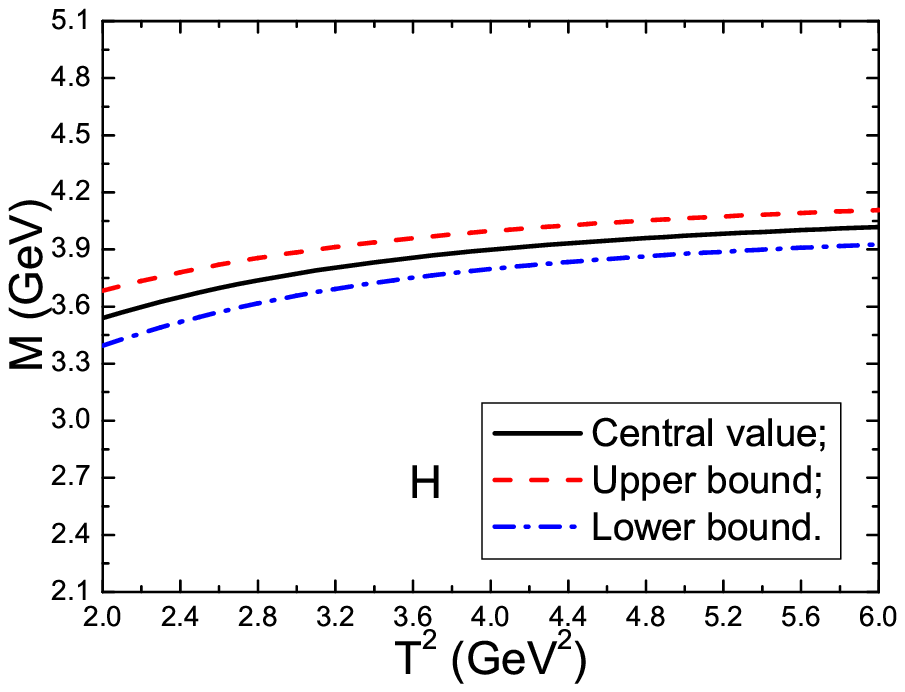}
 \includegraphics[totalheight=3.5cm,width=3.5cm]{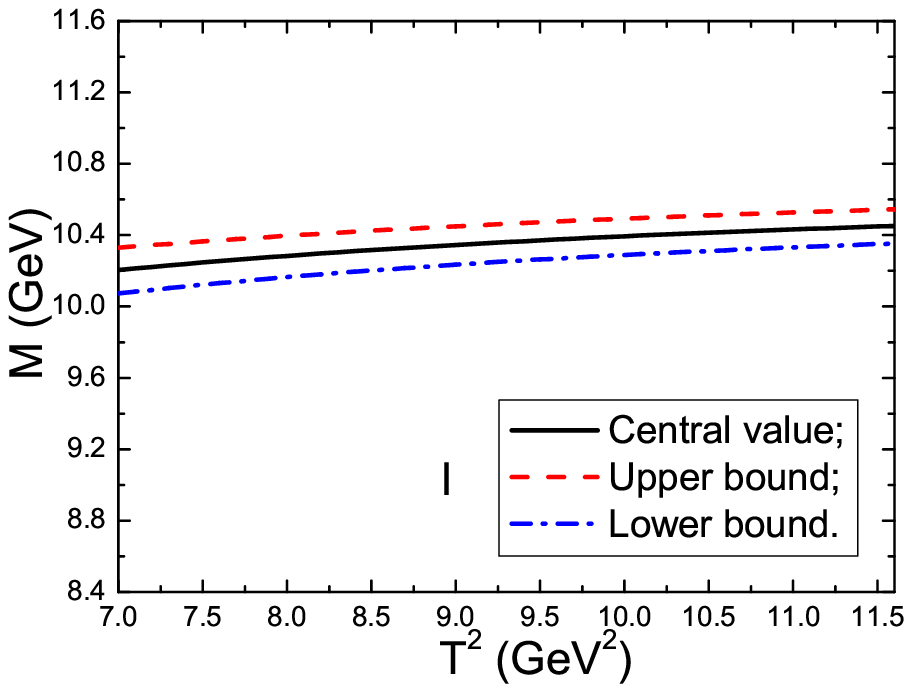}
 \includegraphics[totalheight=3.5cm,width=3.5cm]{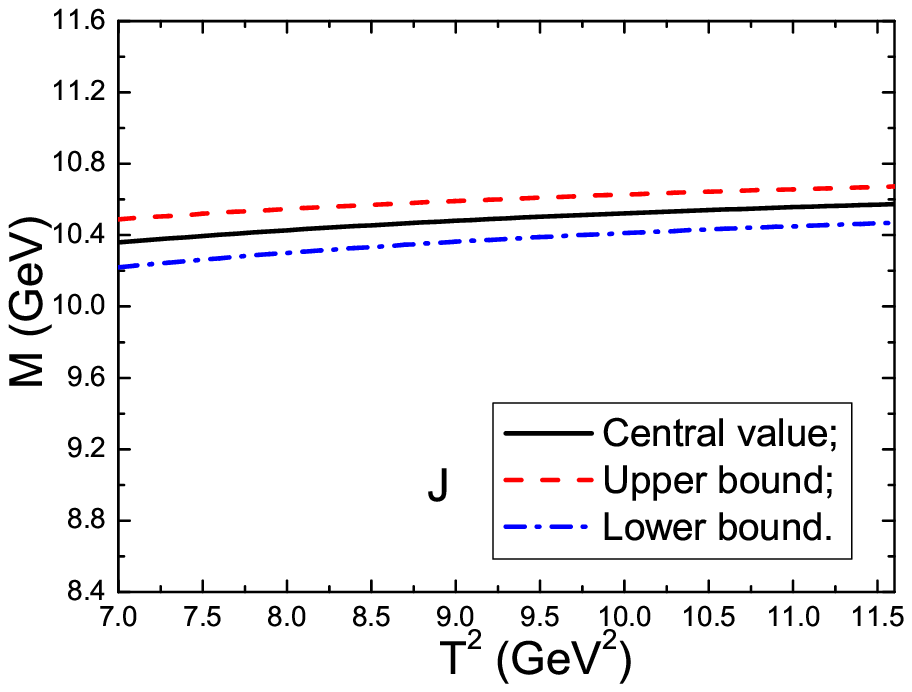}
      \caption{ The  masses of the ${\frac{3}{2}}^-$ heavy and doubly heavy baryon states, the $A$, $B$, $C$, $D$, $E$, $F$, $G$, $H$, $I$ and $J$  correspond
      to the channels $\Sigma_c^*({3\over 2}^-)$, $\Xi_c^*({3\over 2}^-)$, $\Omega_c^*({3\over 2}^-)$, $\Sigma_b^*({3\over 2}^-)$, $\Xi_b^*({3\over 2}^-)$,
 $\Omega_b^*({3\over 2}^-)$, $\Xi^*_{cc}({3\over 2}^-)$, $\Omega^*_{cc}({3\over 2}^-)$, $\Xi^*_{bb}({3\over 2}^-)$ and $\Omega^*_{bb}({3\over 2}^-)$, respectively. }
\end{figure}

\begin{figure}
 \centering
 \includegraphics[totalheight=3.5cm,width=3.5cm]{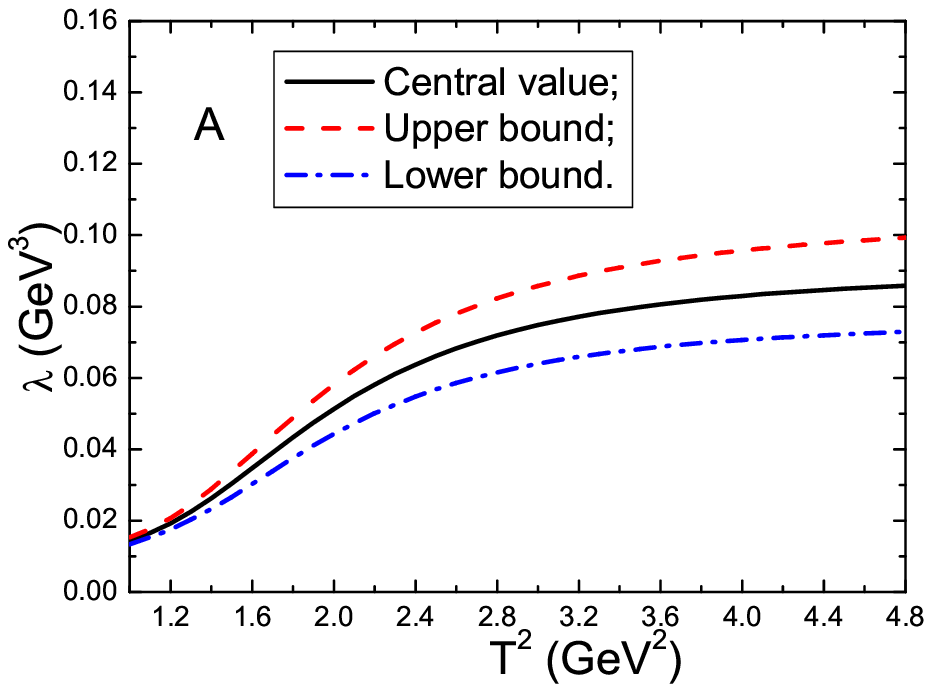}
 \includegraphics[totalheight=3.5cm,width=3.5cm]{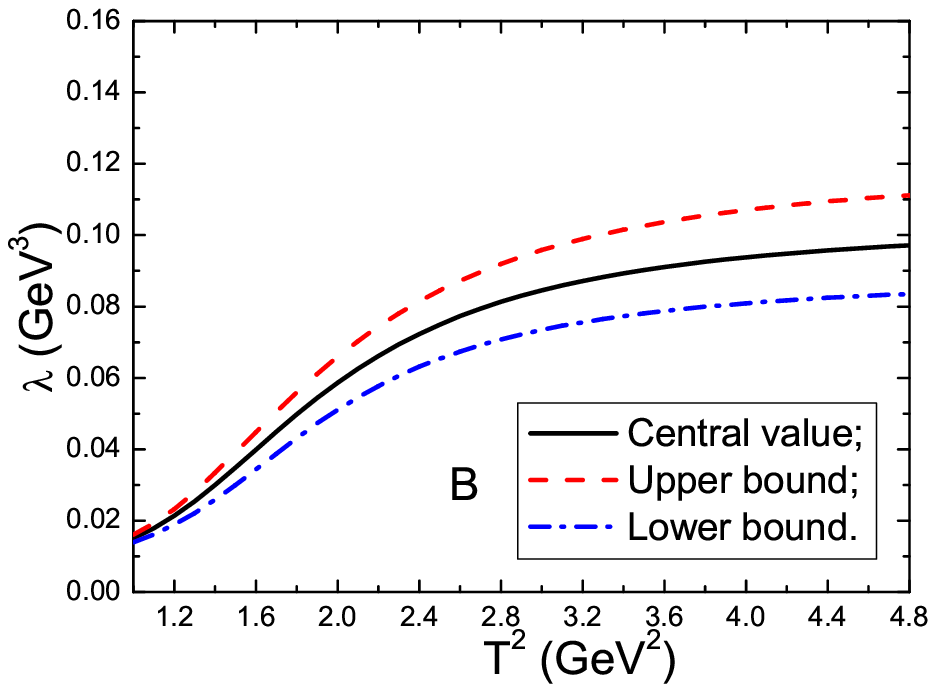}
 \includegraphics[totalheight=3.5cm,width=3.5cm]{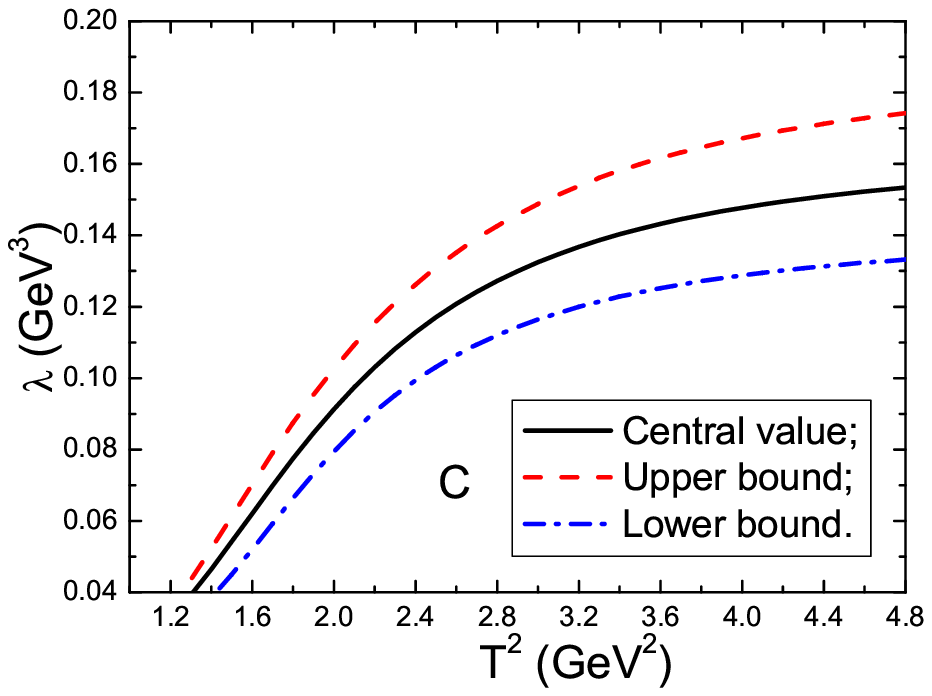}
 \includegraphics[totalheight=3.5cm,width=3.5cm]{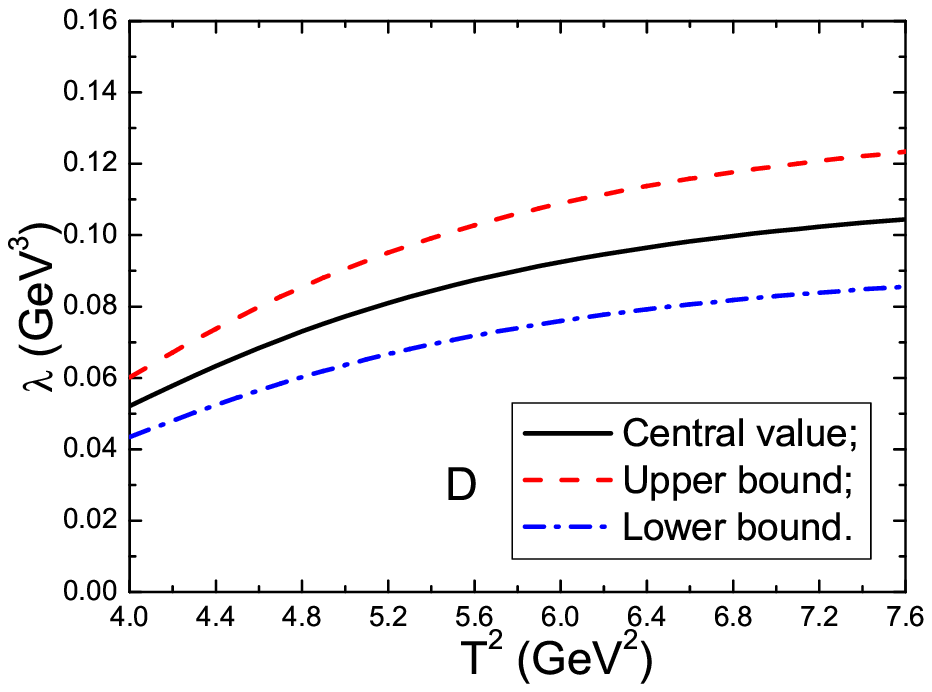}
 \includegraphics[totalheight=3.5cm,width=3.5cm]{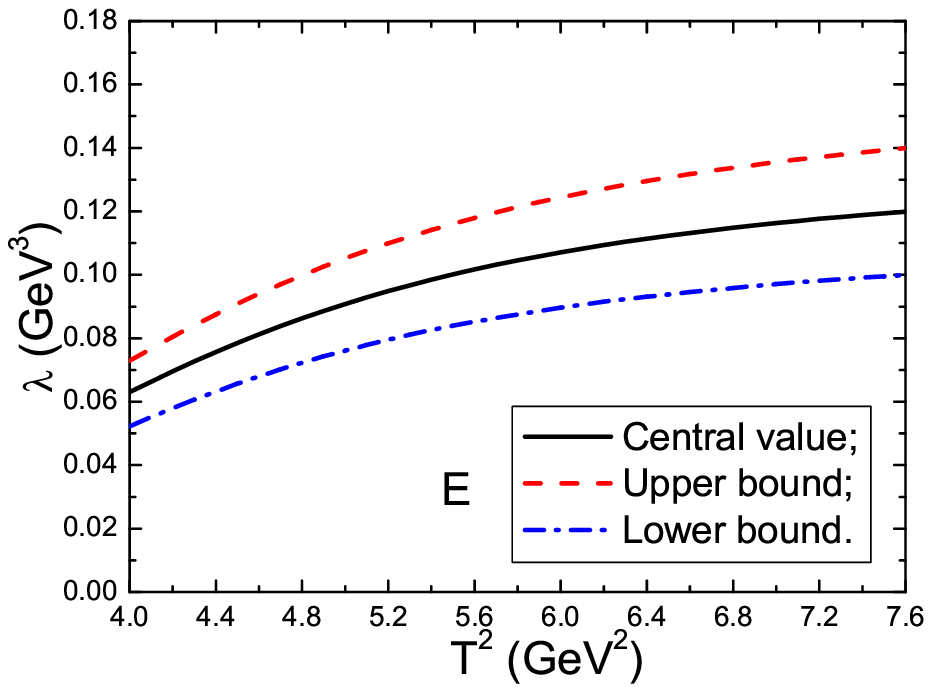}
 \includegraphics[totalheight=3.5cm,width=3.5cm]{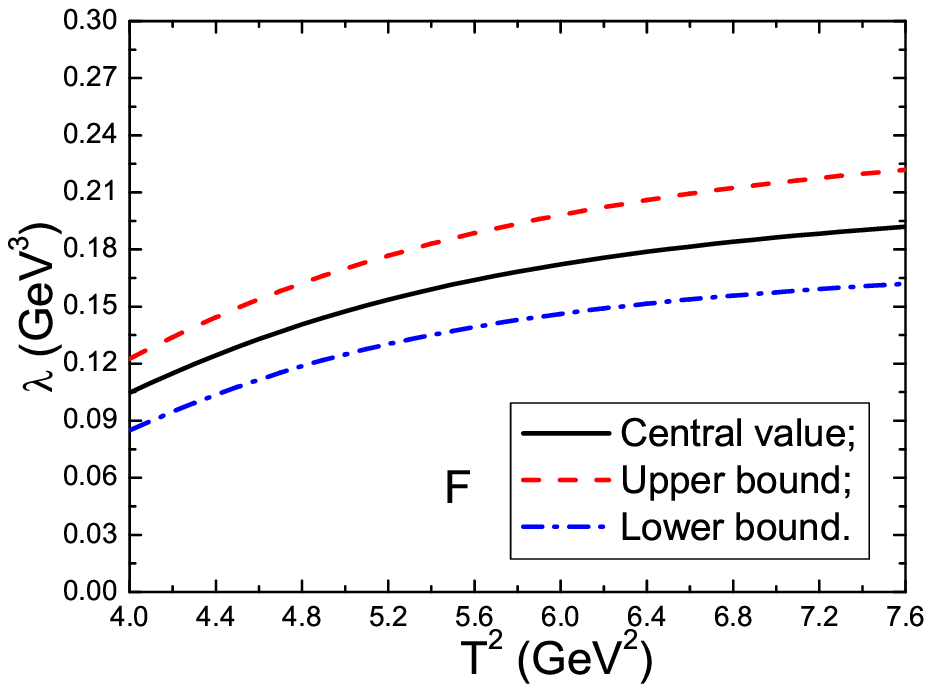}
 \includegraphics[totalheight=3.5cm,width=3.5cm]{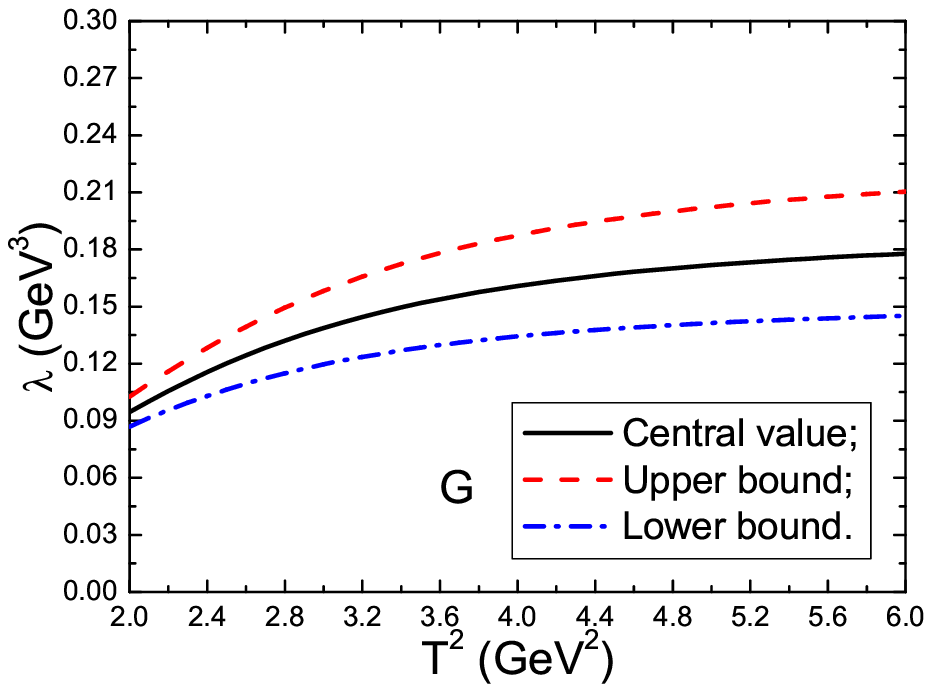}
 \includegraphics[totalheight=3.5cm,width=3.5cm]{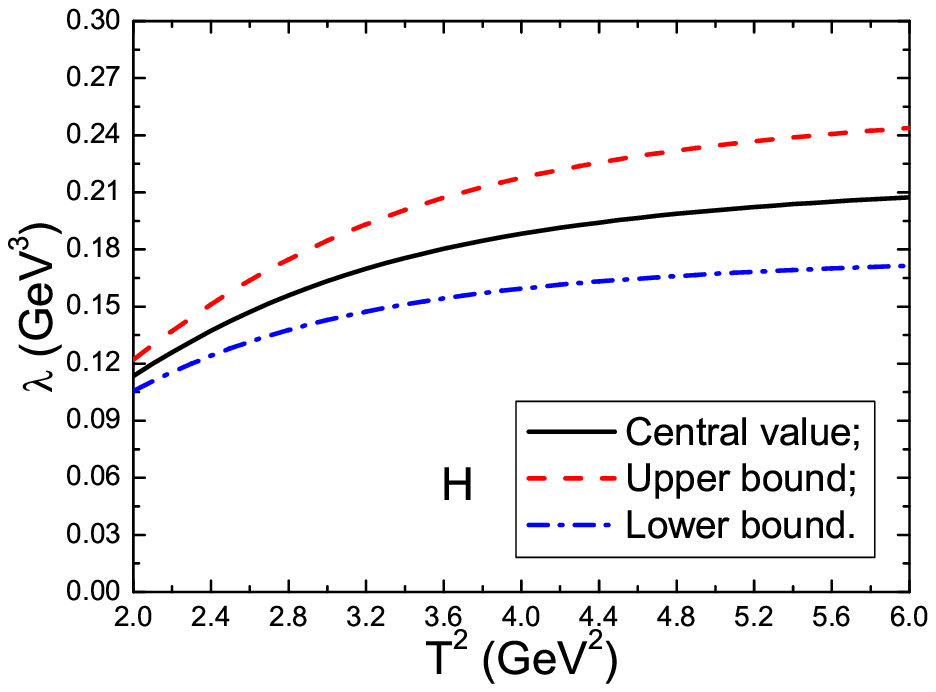}
 \includegraphics[totalheight=3.5cm,width=3.5cm]{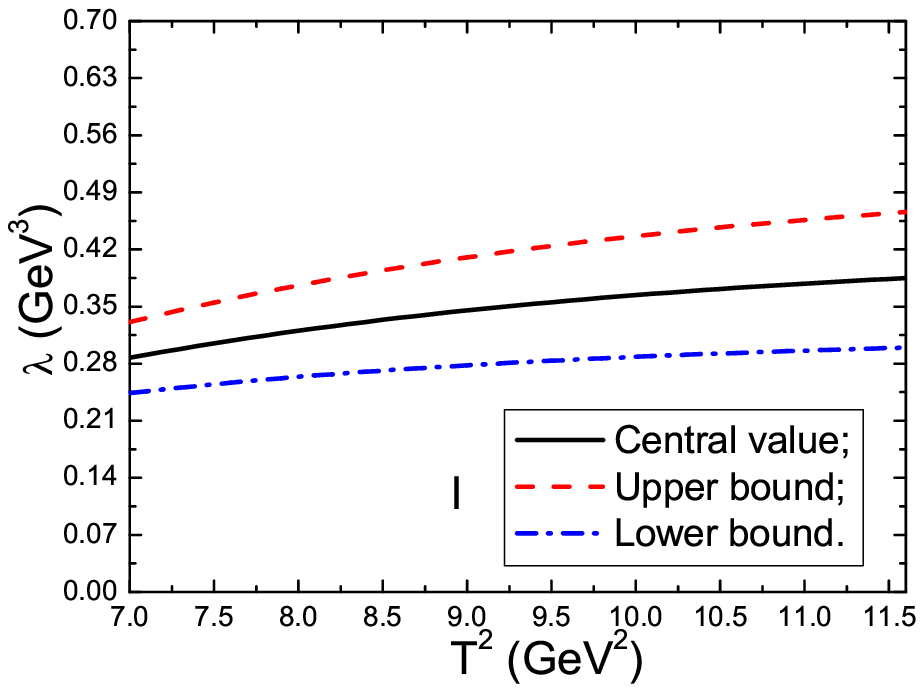}
 \includegraphics[totalheight=3.5cm,width=3.5cm]{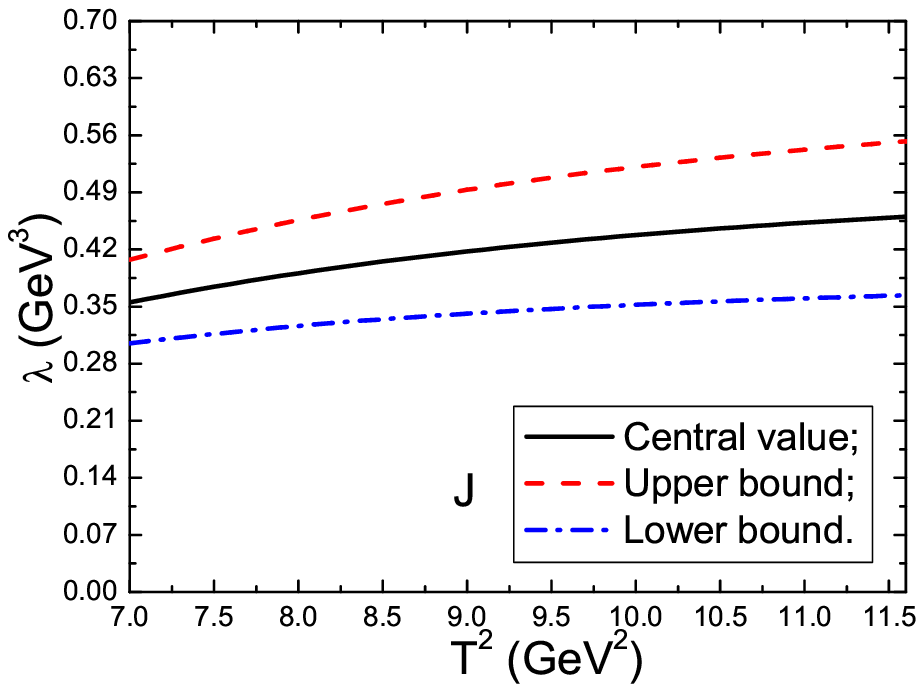}
    \caption{ The  pole residues of the ${\frac{1}{2}}^-$ heavy and doubly heavy baryon states, the $A$, $B$, $C$, $D$, $E$, $F$, $G$, $H$, $I$ and $J$  correspond
      to the channels $\Sigma_c({1\over 2}^-)$, $\Xi'_c({1\over 2}^-)$, $\Omega_c({1\over 2}^-)$, $\Sigma_b({1\over 2}^-)$, $\Xi'_b({1\over 2}^-)$, $\Omega_b({1\over 2}^-)$,
 $\Xi_{cc}({1\over 2}^-)$, $\Omega_{cc}({1\over 2}^-)$, $\Xi_{bb}({1\over 2}^-)$ and $\Omega_{bb}({1\over 2}^-)$, respectively. }
\end{figure}

\begin{figure}
 \centering
 \includegraphics[totalheight=3.5cm,width=3.5cm]{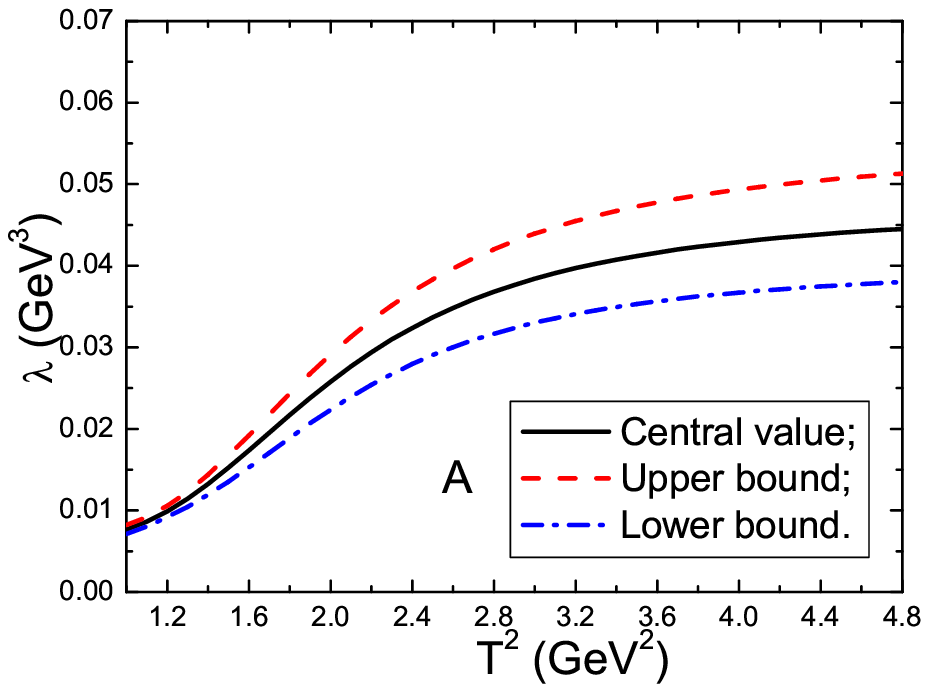}
 \includegraphics[totalheight=3.5cm,width=3.5cm]{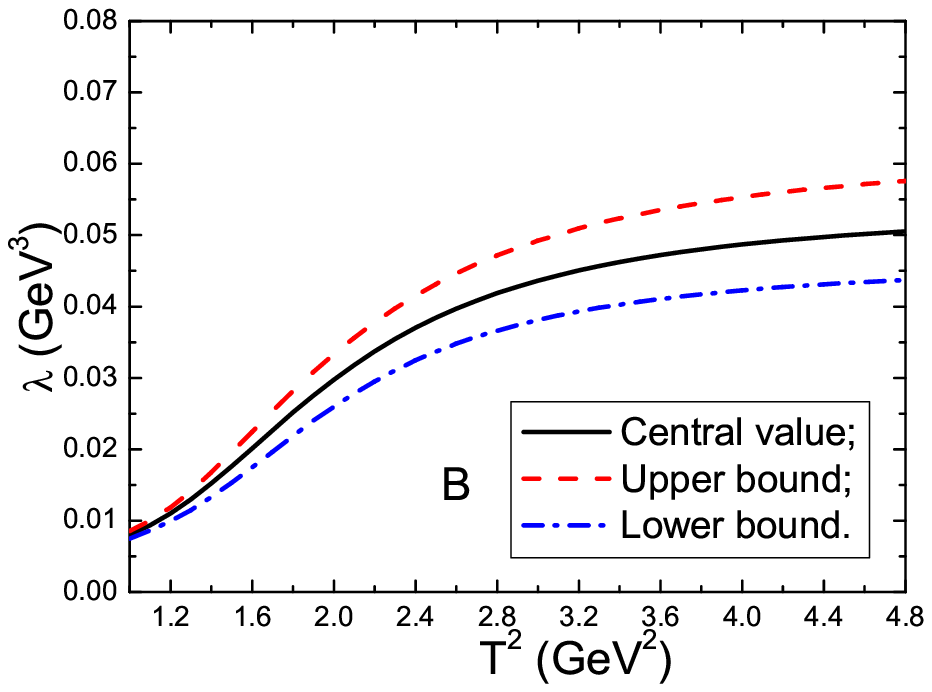}
 \includegraphics[totalheight=3.5cm,width=3.5cm]{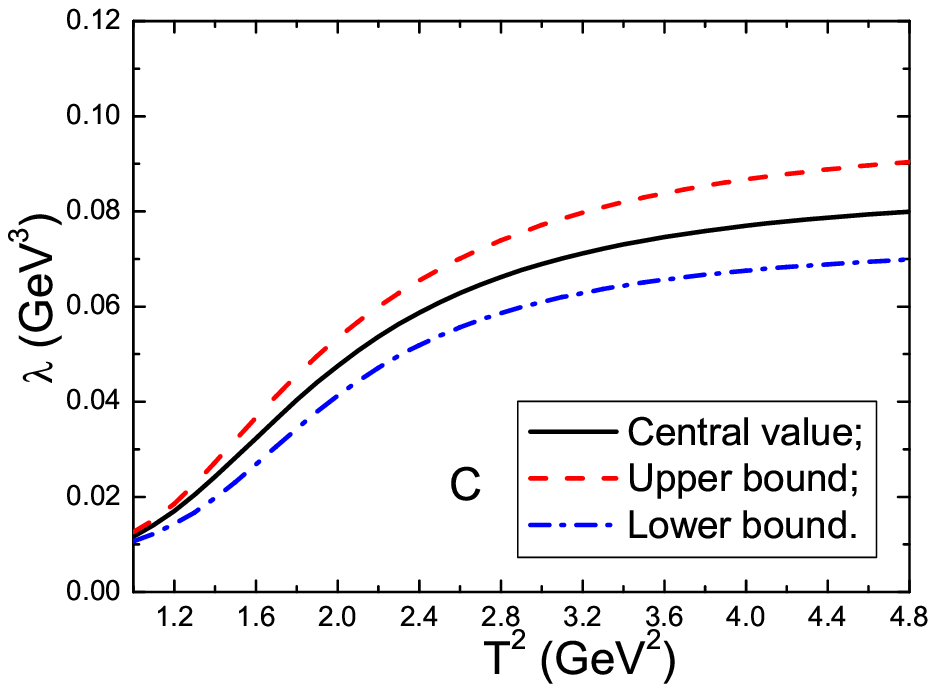}
 \includegraphics[totalheight=3.5cm,width=3.5cm]{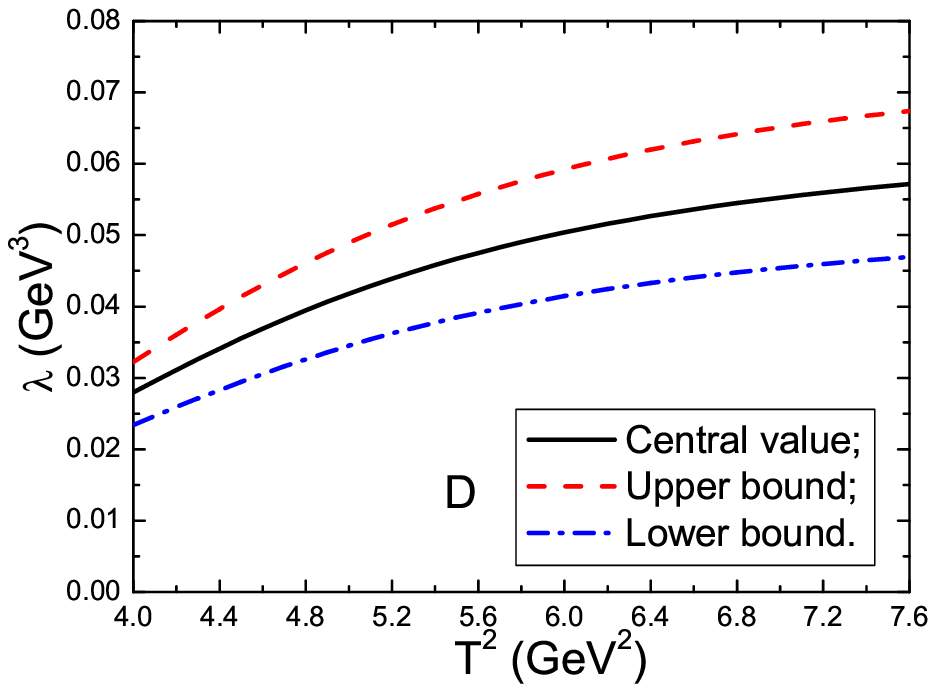}
 \includegraphics[totalheight=3.5cm,width=3.5cm]{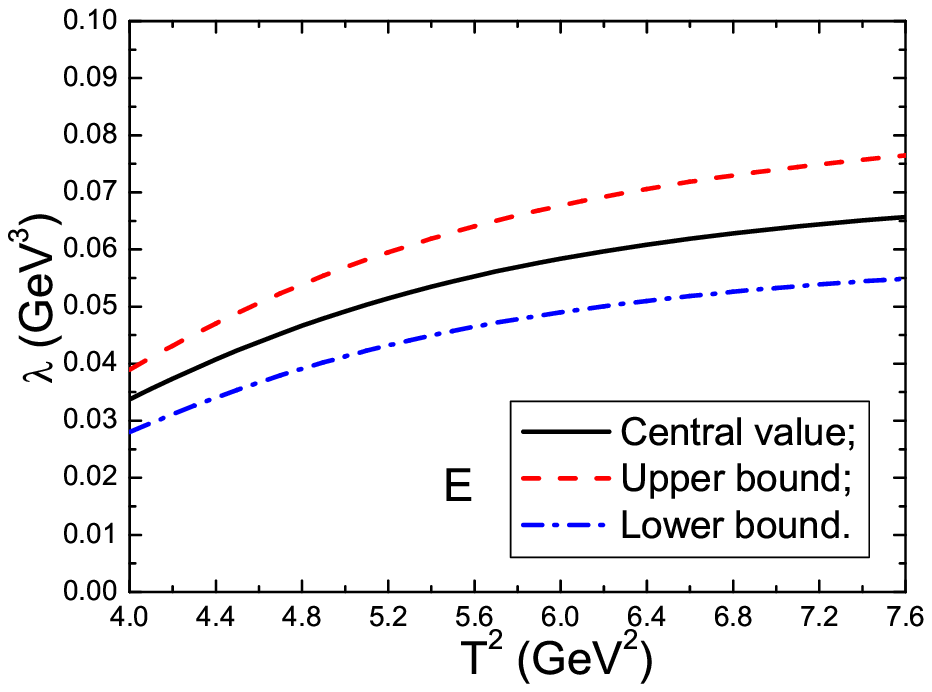}
 \includegraphics[totalheight=3.5cm,width=3.5cm]{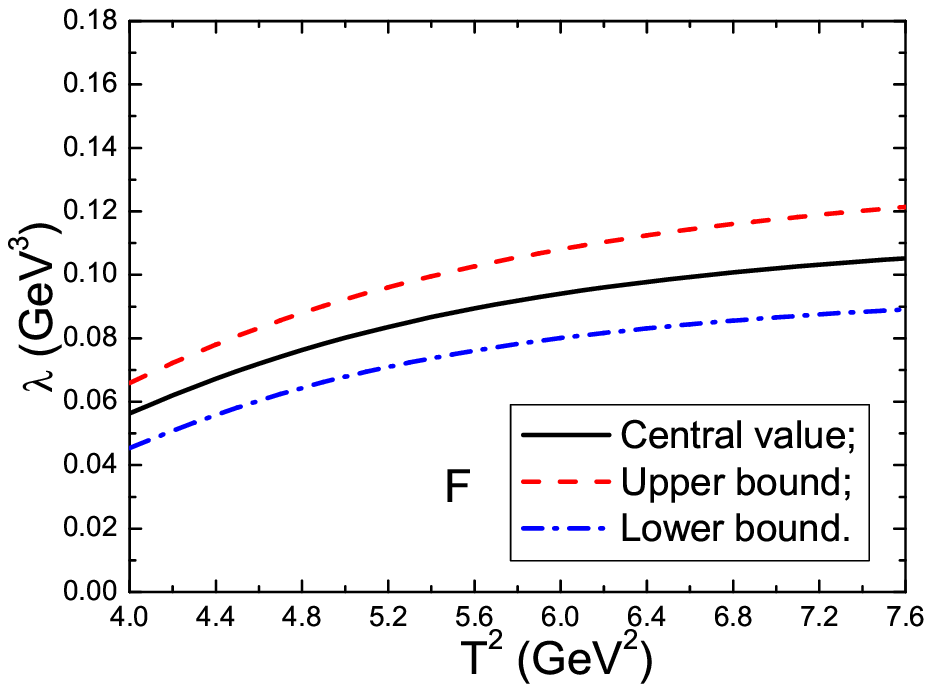}
 \includegraphics[totalheight=3.5cm,width=3.5cm]{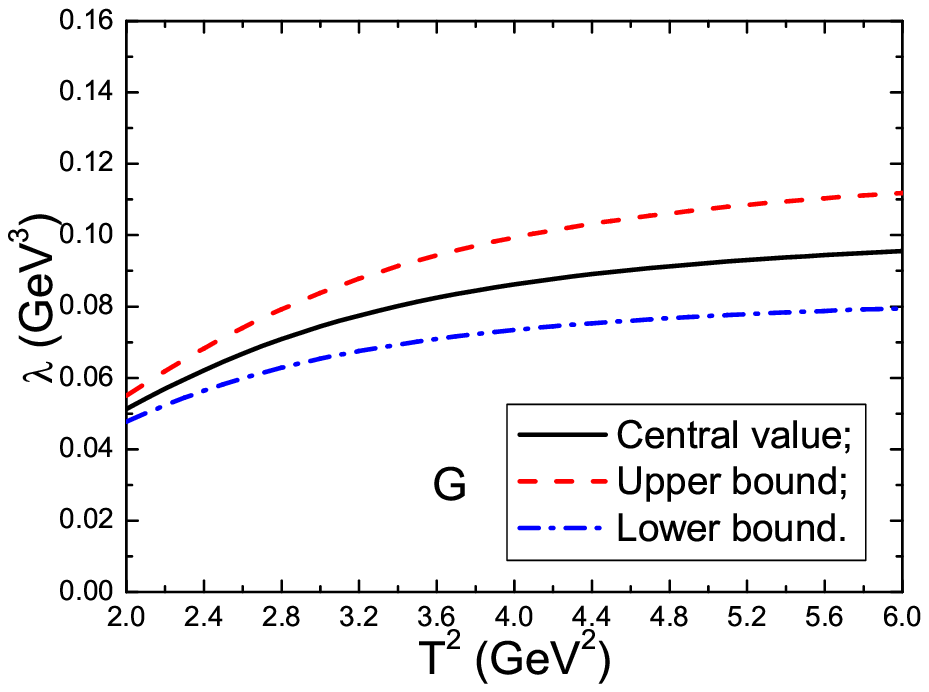}
 \includegraphics[totalheight=3.5cm,width=3.5cm]{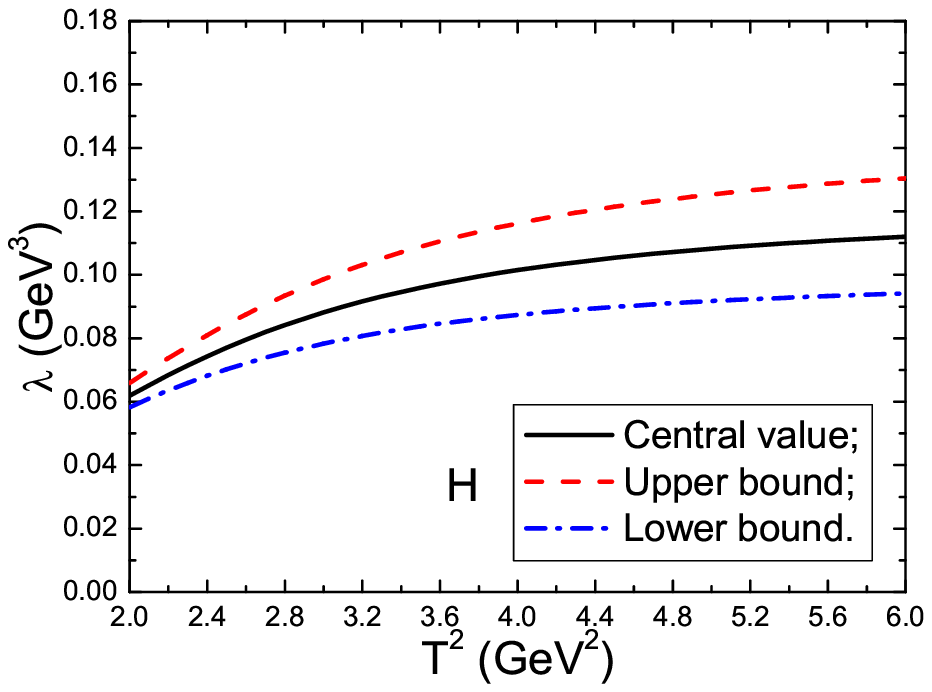}
 \includegraphics[totalheight=3.5cm,width=3.5cm]{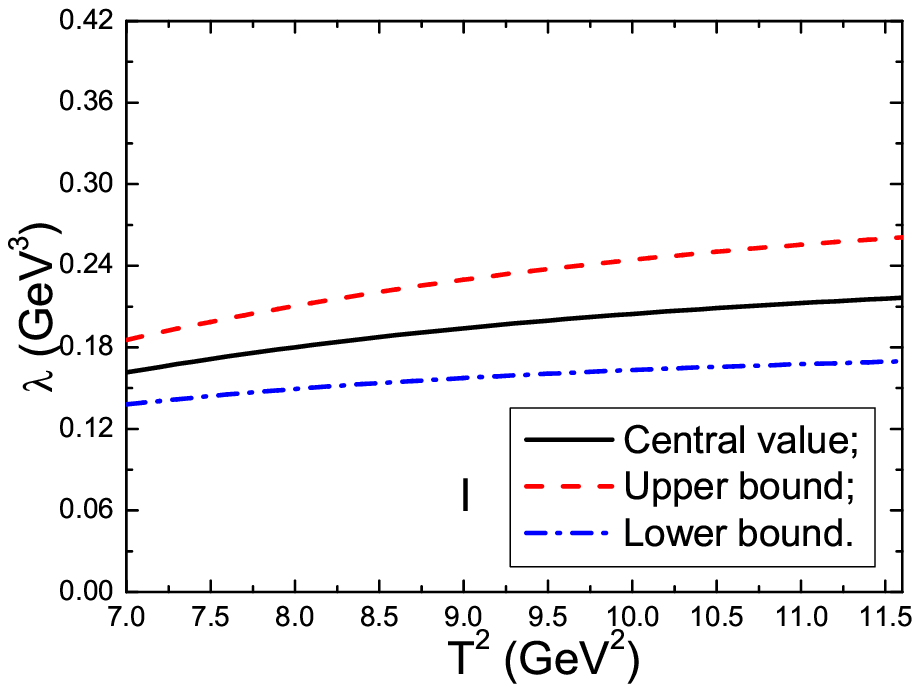}
 \includegraphics[totalheight=3.5cm,width=3.5cm]{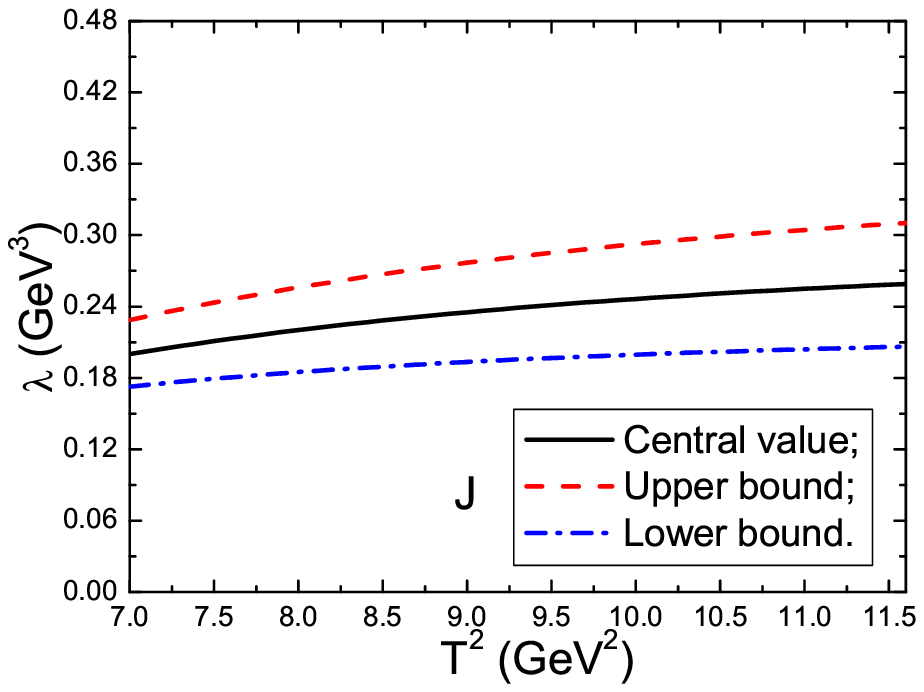}
    \caption{ The  pole residues of the ${\frac{3}{2}}^-$ heavy and doubly heavy baryon states, the $A$, $B$, $C$, $D$, $E$, $F$, $G$, $H$, $I$ and $J$  correspond
      to the channels $\Sigma_c^*({3\over 2}^-)$, $\Xi_c^*({3\over 2}^-)$, $\Omega_c^*({3\over 2}^-)$, $\Sigma_b^*({3\over 2}^-)$, $\Xi_b^*({3\over 2}^-)$,
      $\Omega_b^*({3\over 2}^-)$, $\Xi^*_{cc}({3\over 2}^-)$, $\Omega^*_{cc}({3\over 2}^-)$, $\Xi^*_{bb}({3\over 2}^-)$  and $\Omega^*_{bb}({3\over 2}^-)$, respectively. }
\end{figure}

The charmed partners of the $\Lambda$ baryon  $\Lambda^+_c$,
$\Lambda_c^+(2595)$, $\Lambda_c^+(2625)$, $\Lambda_c^+(2765)$ (or
$\Sigma_c^+(2765)$), $\Lambda_c^+(2880)$
  and $\Lambda_c^+(2940)$ have been observed, the quantum numbers  listed in the  Review of Particle Physics
   are $J^P={\frac{1}{2}}^+$, ${\frac{1}{2}}^-$, ${\frac{3}{2}}^-$,  $?$, ${\frac{5}{2}}^+$ and $?$, respectively
  \cite{PDG}.
  The  flux-tube model favors to identify those charmed baryon states with the spin-parity  ${\frac{1}{2}}^+$, ${\frac{1}{2}}^-$,
  ${\frac{3}{2}}^-$,  ${\frac{3}{2}}^+$, ${\frac{5}{2}}^+$ and ${\frac{5}{2}}^-$, respectively \cite{Zhang2009}.
In Ref.\cite{Wang0101},  we study the mass spectrum  of the ${1\over
2}^\pm$ flavor antitriplet heavy baryon states ($\Lambda_c^+$,
$\Xi_c^+,\Xi_c^0)$ and ($\Lambda_b^0$, $\Xi_b^0,\Xi_b^-)$ by
subtracting the contributions from the corresponding ${1\over
2}^\mp$ heavy baryon states with the QCD sum rules, the predictions
$M_{\Lambda_c}=(2.26 \pm 0.27)\,\rm{GeV}$ and
$M_{\Lambda_c(2595)}=(2.61 \pm 0.21)\,\rm{GeV}$  are in good
agreement with the experimental data.

The $\Xi_c$ baryon states $\Xi_c$, $\Xi^{\prime }_c$, $\Xi_c(2645)$,
$\Xi_c(2790)$, $\Xi_c(2815)$,
 $\Xi_c(2980)$,  $\Xi_c(3055)$, $\Xi_c(3080)$ and $\Xi_c(3123)$
 listed in the Review of Particle Physics have the spin-parity ${\frac{1}{2}}^+$,
  ${\frac{1}{2}}^+$,  ${\frac{3}{2}}^+$,  ${\frac{1}{2}}^-$,  ${\frac{3}{2}}^-$,
  $?$,  $?$,  $?$ and $?$,  respectively \cite{PDG}. The $\Xi_c(2790)$
  and $\Xi_c(2815)$ with the spin-parity ${\frac{1}{2}}^-$ and ${\frac{3}{2}}^-$
  respectively can be interpreted as the charmed-strange analogues of
  the $\Lambda_c^+(2595)$ and $\Lambda_c^+(2625)$, or of the $\Lambda(1405)$ and
$\Lambda(1520)$; they are
  flavor antitriplet  or $\Lambda$-type heavy baryon states
  \cite{HH-Roberts}.  Our
predictions $M_{\Xi_c}=(2.44 \pm 0.23)\,\rm{GeV}$,
$M_{\Xi^{\prime}_c}=(2.56 \pm 0.22)\,\rm{GeV}$,
$M_{\Xi_c(2645)}=(2.65 \pm 0.20)\,\rm{GeV}$ and
$M_{\Xi_c(2790)}=(2.76 \pm 0.18)\,\rm{GeV}$ are in good agreement
with the experimental data \cite{Wang0912,Wang0101,Wang0102}. In the
non-relativistic quark model \cite{HH-Roberts} and the relativistic
quark model based on a quasipotential approach in QCD
\cite{HH-Ebert-1,HH-Ebert-2}, the $\Xi_c(2815)$ is taken as the
$\Lambda$-type baryon state. Although the coupling of an
$\Sigma$-type interpolating  current to a $\Lambda$-type baryon
state is very small, the present prediction
$M_{\Xi_c^*}=(2.86\pm0.17)\,\rm{GeV}$ from the $\Sigma$-type
interpolating  current is compatible with the mass of the
$\Xi_c(2815)$, an $\Sigma$-type $\Xi_c(2815)$ with the spin-parity
${\frac{3}{2}}^-$ is not excluded. The  $\Xi_c(2980)$,
$\Xi_c(3055)$, $\Xi_c(3080)$ and $\Xi_c(3123)$ are unlikely the
${\frac{3}{2}}^-$ state due to their large masses.  The
$\Xi_c(3080)$  is tentatively identified as the strange partner of
the  $\Lambda_c(2880)$ with the spin-parity ${\frac{5}{2}}^+$
\cite{ShortRV2,Cheng2007},  the $\Xi_c(2980)$ maybe have spin-parity
${\frac{1}{2}}^+$ \cite{HH-Ebert-1,Cheng2007}  or ${\frac{3}{2}}^+$
\cite{ShortRV2}, the $\Xi_c(3055)$ and $\Xi_c(3123)$ can be
interpreted  as the second orbital ($1D$) excitations of the $\Xi_c$
with ${\frac{5}{2}}^+$ containing scalar and axial-vector diquarks,
respectively \cite{HH-Ebert-1}.

Only three $\Sigma_c$ baryon states $\Sigma_c(2455)$,
$\Sigma_c(2520)$ and $\Sigma_c(2800)$ have been observed, the
quantum numbers listed in the  Review of Particle Physics are
${\frac{1}{2}}^+$,  ${\frac{3}{2}}^+$  and $?$,  respectively
\cite{PDG}. The possible spin-parity of the $\Sigma_c(2800)$ is
${1\over 2}^-$, ${3\over 2}^-$ and  ${5\over 2}^-$
\cite{HH-Roberts}, or  ${3\over 2}^-$
\cite{ShortRV2,Vijande2007,Cheng2007}. In
  Refs.\cite{Wang0912,Wang0101-2,Wang0102},  we study the ${1\over 2}^+$ and ${3\over 2}^+$ heavy and doubly heavy
  baryon states  by subtracting the contributions from the corresponding
${1\over 2}^-$ and ${3\over 2}^-$ heavy and doubly heavy baryon
states with the QCD sum rules, the predictions
$M_{\Sigma_c(2455)}=(2.40 \pm 0.26)\,\rm{GeV}$ and
$M_{\Sigma_c(2520)}=(2.48 \pm 0.25)\,\rm{GeV}$  are in good
agreement with the experimental data. If we tentatively identify the
$\Sigma_c(2800)$ as the ${3\over 2}^-$ state, the present prediction
$M_{\Sigma_c^*}=(2.74\pm0.20)\,\rm{GeV}$ with the same methods as
Refs.\cite{Wang0912,Wang0101-2,Wang0102} is very good.

The properties of the charmed and doubly charmed baryon states would
be studied at the BESIII and $\rm{\bar{P}ANDA}$ \cite{BESIII,PANDA},
where the charmed baryon states are copiously produced at the
$e^+e^-$ and $p\bar{p}$ collisions. The LHCb is a dedicated $b$ and
$c$-physics precision experiment at the LHC. The LHC will be the
world's most copious  source of the $b$ hadrons, and  a complete
spectrum of the $b$ hadrons will be available through gluon fusion.
In proton-proton collisions at $\sqrt{s}=14\,\rm{TeV}$, the
$b\bar{b}$ cross section is expected to be $\sim 500\mu b$ producing
$10^{12}$ $b\bar{b}$ pairs in a standard  year of running at the
LHCb operational luminosity of $2\times10^{32} \rm{cm}^{-2}
\rm{sec}^{-1}$ \cite{LHC}. The present predictions for the masses of
the ${1\over 2}^-$ and ${3\over 2}^-$ heavy and doubly heavy baryon
states can be confronted with the experimental data  in the future
at the BESIII, $\rm{\bar{P}ANDA}$ and LHCb.

\section{Conclusion}
In this article, we study ${\frac{1}{2}^-}$ and ${\frac{3}{2}^-}$
heavy and doubly heavy baryon states $\Sigma_Q({\frac{1}{2}^-})$,
$\Xi'_Q({\frac{1}{2}^-})$, $\Omega_Q({\frac{1}{2}^-})$,
$\Xi_{QQ}({\frac{1}{2}^-})$, $\Omega_{QQ}({\frac{1}{2}^-})$,
$\Sigma_Q^*({\frac{3}{2}^-})$, $\Xi_Q^*({\frac{3}{2}^-})$,
$\Omega_Q^*({\frac{3}{2}^-})$, $\Xi^*_{QQ}({\frac{3}{2}^-})$ and
$\Omega^*_{QQ}({\frac{3}{2}^-})$ by subtracting the contributions
from the corresponding  ${1\over 2}^+$ and ${3\over 2}^+$ heavy and
doubly heavy  baryon states with the QCD sum rules in a  systematic
way, and make reasonable predictions for their masses.  The present
predictions  can be confronted with the experimental data in the
future at the BESIII, $\rm{\bar{P}ANDA}$ and LHCb, especially the
LHCb. Once reasonable values of the pole residues $\lambda_{-}$ are
obtained, we can take them as   basic input parameters and study the
revelent hadronic processes with the QCD sum rules.

\section*{Acknowledgements}
This  work is supported by National Natural Science Foundation,
Grant Numbers 10775051, 11075053, and Program for New Century
Excellent Talents in University, Grant Number NCET-07-0282, and
Project Supported by Chinese Universities Scientific Fund.

\section*{Appendix}
The spectral densities of the ${\frac{1}{2}^-}$ and
${\frac{3}{2}^-}$ heavy and doubly heavy baryon states
$\Sigma_Q({\frac{1}{2}^-})$, $\Xi'_Q({\frac{1}{2}^-})$,
$\Omega_Q({\frac{1}{2}^-})$, $\Xi_{QQ}({\frac{1}{2}^-})$,
$\Omega_{QQ}({\frac{1}{2}^-})$, $\Sigma_Q^*({\frac{3}{2}^-})$,
$\Xi_Q^*({\frac{3}{2}^-})$, $\Omega_Q^*({\frac{3}{2}^-})$,
$\Xi^*_{QQ}({\frac{3}{2}^-})$ and $\Omega^*_{QQ}({\frac{3}{2}^-})$
at the level of quark-gluon degrees of freedom,

\begin{eqnarray}
\rho^A_{\Sigma_Q}(p_0)&=& \frac{1}{2}\rho^A_{\Omega_Q}(p_0)\mid_{\rm{RP}} \, , \nonumber \\
\rho^B_{\Sigma_Q}(p_0)&=&\frac{1}{2}\rho^B_{\Omega_Q}(p_0)\mid_{\rm{RP}}\,,\nonumber \\
\rho^A_{\Sigma^*_Q}(p_0)&=&\frac{1}{2}\rho^A_{\Omega^*_Q}(p_0)\mid_{\rm{RP}}\,,\nonumber\\
\rho^B_{\Sigma^*_Q}(p_0)&=&\frac{1}{2}\rho^B_{\Omega^*_Q}(p_0)\mid_{\rm{RP}}\,,\nonumber \\
\rho^A_{\Xi_{QQ}}(p_0)&=&\rho^A_{\Omega_{QQ}}(p_0)\mid_{\rm{RP}}\, , \nonumber\\
 \rho^B_{\Xi_{QQ}}(p_0)&=&\rho^B_{\Omega_{QQ}}(p_0)\mid_{\rm{RP}} \, ,\nonumber \\
\rho^A_{\Xi^*_{QQ}}(p_0)&=&\rho^A_{\Omega^*_{QQ}}(p_0)\mid_{\rm{RP}}\, , \nonumber\\
 \rho^B_{\Xi^*_{QQ}}(p_0)&=&\rho^B_{\Omega^*_{QQ}}(p_0)\mid_{\rm{RP}} \, ,
\end{eqnarray}

\begin{eqnarray}
\rho^A_{\Xi'_Q}(p_0)&=&\frac{p_0}{32\pi^4}\int_{t_i}^1dt
t(1-t)^3(p_0^2-\widetilde{m}_Q^2)(5p_0^2-3\widetilde{m}_Q^2)-\frac{p_0m_s\langle\bar{q}q\rangle}{4\pi^2}\int_{t_i}^1
dt t\nonumber\\
&&+\frac{m_s\langle\bar{s}s\rangle}{4\pi^2}\int_{t_i}^1 dt
t(1-t)\left[3p_0+\frac{\widetilde{m}_Q^2}{2}\delta
(p_0-\widetilde{m}_Q)\right]
\nonumber\\
&&-\frac{m_s\langle\bar{s}g_s\sigma Gs\rangle}{24\pi^2}\int_0^1dt
t\left[1+\frac{\widetilde{m}_Q}{4T}\right]\delta
(p_0-\widetilde{m}_Q ) \nonumber\\
&&+\frac{m_s\langle\bar{q}g_s\sigma Gq\rangle}{32\pi^2}\delta
(p_0-m_Q)+\frac{\langle\bar{q}q\rangle\langle\bar{s}s\rangle}{6}\delta(p_0-m_Q)\nonumber \\
&& +\frac{p_0}{96\pi^2}\langle \frac{\alpha_sGG}{\pi}\rangle
\int_{t_i}^1 dt (4-5t)+\frac{1}{192\pi^2}\langle
\frac{\alpha_sGG}{\pi}\rangle \int_{t_i}^1 dt
(1-t)\widetilde{m}_Q^2\delta
(p_0-\widetilde{m}_Q)\nonumber \\
&& -\frac{m_Q^2}{288\pi^2}\langle \frac{\alpha_sGG}{\pi}\rangle
\int_0^1
dt\frac{(1-t)^3}{t^2}\left[1+\frac{\widetilde{m}_Q}{4T}\right]\delta
(p_0-\widetilde{m}_Q) \, ,\\
\rho^B_{\Xi'_Q}(p_0)&=&\frac{3m_Q}{64\pi^4}\int_{t_i}^1dt
(1-t)^2(p_0^2-\widetilde{m}_Q^2)^2-\frac{m_sm_Q\langle\bar{q}q\rangle}{2\pi^2}\int_{t_i}^1
dt+\frac{m_sm_Q\langle\bar{s}s\rangle}{8\pi^2}\int_{t_i}^1 dt
\nonumber\\
&&+\frac{m_s\left[6\langle\bar{q}g_s\sigma
Gq\rangle-\langle\bar{s}g_s\sigma Gs\rangle \right]}{96\pi^2}\delta
(p_0-m_Q) +\frac{\langle\bar{q}q\rangle\langle\bar{s}s\rangle}{3}\delta(p_0-m_Q)\nonumber\\
&&+\frac{m_Q}{192\pi^2}\langle \frac{\alpha_sGG}{\pi}\rangle
\int_{t_i}^1 dt \left[ -3-2t+\frac{2}{t^2}\right]\nonumber \\
&& -\frac{m_Q}{384\pi^2}\langle \frac{\alpha_sGG}{\pi}\rangle
\int_0^1 dt\frac{(1-t)^2}{t}\widetilde{m}_Q\delta
(p_0-\widetilde{m}_Q) \, ,
\end{eqnarray}

\begin{eqnarray}
\rho^A_{\Omega_Q}(p_0)&=&\frac{p_0}{16\pi^4}\int_{t_i}^1dt
t(1-t)^3(p_0^2-\widetilde{m}_Q^2)(5p_0^2-3\widetilde{m}_Q^2)-\frac{p_0m_s\langle\bar{s}s\rangle}{\pi^2}\int_{t_i}^1
dt t\nonumber\\
&&+\frac{m_s\langle\bar{s}s\rangle}{\pi^2}\int_{t_i}^1 dt
t(1-t)\left[3p_0+\frac{\widetilde{m}_Q^2}{2}\delta
(p_0-\widetilde{m}_Q)\right]
\nonumber\\
&&-\frac{m_s\langle\bar{s}g_s\sigma Gs\rangle}{6\pi^2}\int_0^1dt
t\left[1+\frac{\widetilde{m}_Q}{4T}\right]\delta
(p_0-\widetilde{m}_Q) +\frac{\langle\bar{s}s\rangle^2}{3}\delta(p_0-m_Q)\nonumber\\
&&+\frac{m_s\langle\bar{s}g_s\sigma Gs\rangle}{8\pi^2}\delta
(p_0-m_Q)+\frac{p_0}{48\pi^2}\langle \frac{\alpha_sGG}{\pi}\rangle
\int_{t_i}^1 dt (4-5t)\nonumber \\
&&+\frac{1}{96\pi^2}\langle \frac{\alpha_sGG}{\pi}\rangle
\int_{t_i}^1 dt (1-t)\widetilde{m}_Q^2\delta
(p_0-\widetilde{m}_Q)\nonumber \\
&& -\frac{m_Q^2}{144\pi^2}\langle \frac{\alpha_sGG}{\pi}\rangle
\int_0^1 dt\frac{(1-t)^3}{t^2}\left[1+\frac{p_0}{4T}\right]\delta
(p_0-\widetilde{m}_Q) \, ,\\
\rho^B_{\Omega_Q}(p_0)&=&\frac{3m_Q}{32\pi^4}\int_{t_i}^1dt
(1-t)^2(p_0^2-\widetilde{m}_Q^2)^2-\frac{3m_sm_Q\langle\bar{s}s\rangle}{2\pi^2}\int_{t_i}^1
dt\nonumber\\
&&+\frac{5m_s \langle\bar{s}g_s\sigma Gs\rangle }{24\pi^2}\delta
(p_0-m_Q) +\frac{2\langle\bar{s}s\rangle^2}{3}\delta(p_0-m_Q)\nonumber\\
&&+\frac{m_Q}{96\pi^2}\langle \frac{\alpha_sGG}{\pi}\rangle
\int_{t_i}^1 dt \left[ -3-2t+\frac{2}{t^2}\right]\nonumber \\
&& -\frac{m_Q}{192\pi^2}\langle \frac{\alpha_sGG}{\pi}\rangle
\int_0^1 dt\frac{(1-t)^2}{t}\widetilde{m}_Q \delta
(p_0-\widetilde{m}_Q) \, ,
\end{eqnarray}

\begin{eqnarray}
\rho^A_{\Omega_{QQ}}(p_0)&=&\frac{3p_0}{8 \pi^4}
\int_{\alpha_{i}}^{\alpha_{f}}d\alpha \int_{\beta_{i}}^{1-\alpha}
d\beta\alpha\beta(1-\alpha-\beta)(p_0^2-\widetilde{m}^2_Q)(5p_0^2-3\widetilde{m}^2_Q)
\nonumber\\
&&+\frac{3m_Q^2p_0}{8\pi^4}\int_{\alpha_{i}}^{\alpha_{f}}d\alpha
\int_{\beta_{i}}^{1-\alpha} d\beta
(1-\alpha-\beta)(p_0^2-\widetilde{m}^2_Q) \nonumber\\
&&-\frac{m_Q^2}{24\pi^2}
\langle\frac{\alpha_sGG}{\pi}\rangle\int_{\alpha_{i}}^{\alpha_{f}}d\alpha
\int_{\beta_{i}}^{1-\alpha} d\beta (1-\alpha-\beta)
\left[\frac{\alpha}{\beta^2}+\frac{\beta}{\alpha^2} \right]\left[1+\frac{p_0}{4T}\right]\delta(p_0-\widetilde{m}_Q)\nonumber\\
&&-\frac{m_Q^4}{192\pi^2p_0T}\langle\frac{\alpha_sGG}{\pi}\rangle\int_{\alpha_{i}}^{\alpha_{f}}d\alpha
\int_{\beta_{i}}^{1-\alpha} d\beta (1-\alpha-\beta)\left[\frac{1}{\alpha^3}+\frac{1}{\beta^3}\right]\delta(p_0-\widetilde{m}_Q)\nonumber\\
&&+\frac{m_Q^2}{32\pi^2}\langle\frac{\alpha_sGG}{\pi}\rangle\int_{\alpha_{i}}^{\alpha_{f}}d\alpha
\int_{\beta_{i}}^{1-\alpha} d\beta (1-\alpha-\beta)\left[\frac{1}{\alpha^2}+\frac{1}{\beta^2}\right]\delta(p_0-\widetilde{m}_Q)\nonumber\\
&&+\frac{m_s\langle\bar{s}{s}\rangle}{4\pi^2}\int_{\alpha_{i}}^{\alpha_{f}}d\alpha
\alpha(1-\alpha)\left[
6p_0+p_0^2\delta(p_0-\widetilde{\widetilde{m}}_Q)\right]
+\frac{m_sm_Q^2\langle\bar{s}{s}\rangle}{8\pi^2}\int_{\alpha_{i}}^{\alpha_{f}}d\alpha
\delta(p_0-\widetilde{\widetilde{m}}_Q)  \nonumber \\
&&+\frac{1}{32\pi^2}\langle\frac{\alpha_sGG}{\pi}\rangle\int_{\alpha_{i}}^{\alpha_{f}}d\alpha
\int_{\beta_{i}}^{1-\alpha} d\beta (\alpha+\beta)
\left[3p_0+\frac{\widetilde{m}_Q^2}{2}\delta(p_0-\widetilde{m}_Q)
\right]\nonumber \\
&&+\frac{m_Q^2}{64\pi^2}\langle\frac{\alpha_sGG}{\pi}\rangle\int_{\alpha_{i}}^{\alpha_{f}}d\alpha
\int_{\beta_{i}}^{1-\alpha} d\beta \frac{\alpha+\beta}{\alpha\beta}
 \delta(p_0-\widetilde{m}_Q) \nonumber\\
 &&-\frac{m_s\langle\bar{s}g_s\sigma G{s}\rangle}{4\pi^2}\int_{\alpha_{i}}^{\alpha_{f}}d\alpha
\alpha(1-\alpha)\left[1+\frac{3p_0}{8T}+\frac{p_0^2}{16T^2}\right]
\delta(p_0-\widetilde{\widetilde{m}}_Q)\, ,
\end{eqnarray}
\begin{eqnarray}
\rho^B_{\Omega_{QQ}}(p_0)&=&\frac{3m_s}{8 \pi^4}
\int_{\alpha_{i}}^{\alpha_{f}}d\alpha \int_{\beta_{i}}^{1-\alpha}
d\beta\alpha\beta(p_0^2-\widetilde{m}^2_Q)(2p_0^2-\widetilde{m}^2_Q)
\nonumber\\
&&+\frac{3m_sm_Q^2}{4\pi^4}\int_{\alpha_{i}}^{\alpha_{f}}d\alpha
\int_{\beta_{i}}^{1-\alpha} d\beta
(p_0^2-\widetilde{m}^2_Q) \nonumber\\
&&-\frac{m_sm_Q^2}{96\pi^2}
\langle\frac{\alpha_sGG}{\pi}\rangle\int_{\alpha_{i}}^{\alpha_{f}}d\alpha
\int_{\beta_{i}}^{1-\alpha} d\beta
\left[\frac{\alpha}{\beta^2}+\frac{\beta}{\alpha^2} \right]\left[\frac{1}{\widetilde{m}_Q}+\frac{1}{2T}\right]\delta(p_0-\widetilde{m}_Q)\nonumber\\
&&-\frac{m_sm_Q^4}{96\pi^2p_0^2T}\langle\frac{\alpha_sGG}{\pi}\rangle\int_{\alpha_{i}}^{\alpha_{f}}d\alpha
\int_{\beta_{i}}^{1-\alpha} d\beta \left[\frac{1}{\alpha^3}+\frac{1}{\beta^3}\right]\delta(p_0-\widetilde{m}_Q)\nonumber\\
&&+\frac{m_sm_Q^2}{16\pi^2p_0}\langle\frac{\alpha_sGG}{\pi}\rangle\int_{\alpha_{i}}^{\alpha_{f}}d\alpha
\int_{\beta_{i}}^{1-\alpha} d\beta \left[\frac{1}{\alpha^2}+\frac{1}{\beta^2}\right]\delta(p_0-\widetilde{m}_Q)\nonumber\\
&&-\frac{\langle\bar{s}{s}\rangle}{2\pi^2}\int_{\alpha_{i}}^{\alpha_{f}}d\alpha
\alpha(1-\alpha)\left[ 3p_0^2-2\widetilde{\widetilde{m}}_Q^2\right]
-\frac{m_Q^2\langle\bar{s}{s}\rangle}{\pi^2}\int_{\alpha_{i}}^{\alpha_{f}}d\alpha
 \nonumber \\
&&-\frac{m_s}{16\pi^2}\langle\frac{\alpha_sGG}{\pi}\rangle\int_{\alpha_{i}}^{\alpha_{f}}d\alpha
\int_{\beta_{i}}^{1-\alpha} d\beta
\left[1+\frac{\widetilde{m}_Q}{4}\delta(p_0-\widetilde{m}_Q)\right]\nonumber\\
&&+\frac{3\langle\bar{s}g_s\sigma
G{s}\rangle}{4\pi^2}\int_{\alpha_{i}}^{\alpha_{f}}d\alpha
\alpha(1-\alpha)\left[1+\left(\frac{p_0}{2}+\frac{p_0^2}{8T}\right)
\delta(p_0-\widetilde{\widetilde{m}}_Q)\right]\nonumber\\
&&-\frac{\langle\bar{s}g_s\sigma
G{s}\rangle}{8\pi^2}\int_{\alpha_{i}}^{\alpha_{f}}d\alpha
\alpha(1-\alpha)\left[1+\frac{3p_0}{4}
\delta(p_0-\widetilde{\widetilde{m}}_Q)\right]\,,
\end{eqnarray}

\begin{eqnarray}
\rho^A_{\Xi^*_Q}(p_0)&=&\frac{p_0}{128\pi^4}\int_{t_i}^1dt
t(2+t)(1-t)^2(p_0^2-\widetilde{m}_Q^2)^2+\frac{p_0m_s\langle\bar{s}s\rangle}{16\pi^2}\int_{t_i}^1
dt t^2\nonumber\\
&&-\frac{m_s\langle\bar{q}q\rangle}{8\pi^2}\int_{t_i}^1 dt t
+\frac{m_s\langle\bar{s}g_s\sigma Gs\rangle}{192\pi^2}\int_0^1dt t
\delta (p_0-\widetilde{m}_Q ) \nonumber\\
&&+\frac{m_s \left[3\langle\bar{q}g_s\sigma
Gq\rangle-\langle\bar{s}g_s\sigma Gs\rangle \right]}{192\pi^2}\delta
(p_0-m_Q)+\frac{\langle\bar{q}q\rangle\langle\bar{s}s\rangle}{12}\delta(p_0-m_Q)\nonumber \\
&& -\frac{p_0}{384\pi^2}\langle \frac{\alpha_sGG}{\pi}\rangle
\int_{t_i}^1 dt t(2-t) -\frac{m_Q^2}{2304\pi^2}\langle
\frac{\alpha_sGG}{\pi}\rangle \int_{t_i}^1 dt
\frac{t^3-3t+2}{t^2}\delta (p_0-\widetilde{m}_Q) \,, \\
\rho^B_{\Xi^*_Q}(p_0)&=&\frac{m_Q}{128\pi^4}\int_{t_i}^1dt
(2+t)(1-t)^2(p_0^2-\widetilde{m}_Q^2)^2+\frac{m_sm_Q\langle\bar{s}s\rangle}{16\pi^2}\int_{t_i}^1
dt t- \frac{m_sm_Q\langle\bar{q}q\rangle}{8\pi^2}\int_{t_i}^1 dt
\nonumber\\
&&+\frac{m_sm_Q\langle\bar{s}g_s\sigma
Gs\rangle}{192\pi^2p_0}\int_0^1dt t \delta (p_0-\widetilde{m}_Q
)+\frac{m_s\left[3\langle\bar{q}g_s\sigma
Gq\rangle-\langle\bar{s}g_s\sigma Gs\rangle \right]}{192\pi^2}\delta
(p_0-m_Q)\nonumber\\
 && +\frac{\langle\bar{q}q\rangle\langle\bar{s}s\rangle}{12}\delta(p_0-m_Q)-\frac{m_Q}{384\pi^2}\langle \frac{\alpha_sGG}{\pi}\rangle
\int_{t_i}^1 dt (2-t)\nonumber\\
&&+\frac{m_Q}{1152\pi^2}\langle \frac{\alpha_sGG}{\pi}\rangle
\int_{t_i}^1 dt \frac{(1-t)^3(3t+4)}{t^2}\nonumber \\
&& -\frac{m_Q}{2304\pi^2}\langle \frac{\alpha_sGG}{\pi}\rangle
\int_0^1 dt\frac{t^3-3t+2}{t}\widetilde{m}_Q \delta
(p_0-\widetilde{m}_Q) \, ,
\end{eqnarray}

\begin{eqnarray}
\rho^A_{\Omega^*_Q}(p_0)&=&\frac{p_0}{64\pi^4}\int_{t_i}^1dt
t(2+t)(1-t)^2(p_0^2-\widetilde{m}_Q^2)^2-\frac{p_0m_s\langle\bar{s}s\rangle}{4\pi^2}\int_{t_i}^1
dt t(2-t)\nonumber\\
&&+\frac{m_s\langle\bar{s}g_s\sigma Gs\rangle}{48\pi^2}\int_0^1dt
t\delta (p_0-\widetilde{m}_Q) +\frac{m_s\langle\bar{s}g_s\sigma
Gs\rangle}{24\pi^2}\delta
(p_0-m_Q)+\frac{\langle\bar{s}s\rangle^2}{6}\delta(p_0-m_Q)\nonumber \\
&&-\frac{p_0}{192\pi^2}\langle \frac{\alpha_sGG}{\pi}\rangle
\int_{t_i}^1 dt t(2-t) +\frac{m_Q^2}{1152\pi^2}\langle
\frac{\alpha_sGG}{\pi}\rangle \int_{t_i}^1 dt
\frac{(1-t)^3}{t^2}\delta (p_0-\widetilde{m}_Q)
\nonumber \\
&& -\frac{m_Q^2}{384\pi^2}\langle \frac{\alpha_sGG}{\pi}\rangle
\int_0^1 dt\frac{(1-t)^2}{t^2}\delta (p_0-\widetilde{m}_Q) \, ,
\\
\rho^B_{\Omega^*_Q}(p_0)&=&\frac{m_Q}{64\pi^4}\int_{t_i}^1dt
(2+t)(1-t)^2(p_0^2-\widetilde{m}_Q^2)^2-\frac{m_sm_Q\langle\bar{s}s\rangle}{4\pi^2}\int_{t_i}^1
dt (2-t)
\nonumber\\
&&+\frac{m_sm_Q\langle\bar{s}g_s\sigma
Gs\rangle}{48\pi^2p_0}\int_0^1dt \delta
(p_0-\widetilde{m}_Q)+\frac{m_s \langle\bar{s}g_s\sigma Gs\rangle
}{24\pi^2}\delta
(p_0-m_Q) +\frac{\langle\bar{s}s\rangle^2}{6}\delta(p_0-m_Q)\nonumber\\
&&-\frac{m_Q}{192\pi^2}\langle \frac{\alpha_sGG}{\pi}\rangle
\int_{t_i}^1 dt (2-t)+\frac{m_Q}{576\pi^2}\langle
\frac{\alpha_sGG}{\pi}\rangle
\int_{t_i}^1 dt \frac{(1-t)^3(3t+4)}{t^2}\nonumber \\
&& -\frac{m_Q}{1152\pi^2}\langle \frac{\alpha_sGG}{\pi}\rangle
\int_0^1 dt\frac{t^3-3t+2}{t}\widetilde{m}_Q \delta
(p_0-\widetilde{m}_Q) \, ,
\end{eqnarray}

\begin{eqnarray}
\rho^A_{\Omega^*_{QQ}}(p_0)&=&\frac{3p_0}{16 \pi^4}
\int_{\alpha_{i}}^{\alpha_{f}}d\alpha \int_{\beta_{i}}^{1-\alpha}
d\beta\alpha\beta(1-\alpha-\beta)(p_0^2-\widetilde{m}^2_Q)(2p_0^2-\widetilde{m}^2_Q)
\nonumber\\
&&+\frac{3m_Q^2p_0}{16\pi^4}\int_{\alpha_{i}}^{\alpha_{f}}d\alpha
\int_{\beta_{i}}^{1-\alpha} d\beta
(1-\alpha-\beta)(p_0^2-\widetilde{m}^2_Q) \nonumber\\
&&-\frac{m_Q^2}{192\pi^2}
\langle\frac{\alpha_sGG}{\pi}\rangle\int_{\alpha_{i}}^{\alpha_{f}}d\alpha
\int_{\beta_{i}}^{1-\alpha} d\beta (1-\alpha-\beta)
\left[\frac{\alpha}{\beta^2}+\frac{\beta}{\alpha^2} \right]\left[1+\frac{p_0}{2T}\right]\delta(p_0-\widetilde{m}_Q)\nonumber\\
&&-\frac{m_Q^4}{384\pi^2p_0T}\langle\frac{\alpha_sGG}{\pi}\rangle\int_{\alpha_{i}}^{\alpha_{f}}d\alpha
\int_{\beta_{i}}^{1-\alpha} d\beta (1-\alpha-\beta)\left[\frac{1}{\alpha^3}+\frac{1}{\beta^3}\right]\delta(p_0-\widetilde{m}_Q)\nonumber\\
&&+\frac{m_Q^2}{64\pi^2}\langle\frac{\alpha_sGG}{\pi}\rangle\int_{\alpha_{i}}^{\alpha_{f}}d\alpha
\int_{\beta_{i}}^{1-\alpha} d\beta (1-\alpha-\beta)\left[\frac{1}{\alpha^2}+\frac{1}{\beta^2}\right]\delta(p_0-\widetilde{m}_Q)\nonumber\\
&&+\frac{m_s\langle\bar{s}{s}\rangle}{4\pi^2}\int_{\alpha_{i}}^{\alpha_{f}}d\alpha
\alpha(1-\alpha)\left[
p_0+\frac{p_0^2}{4}\delta(p_0-\widetilde{\widetilde{m}}_Q)\right]
+\frac{m_sm_Q^2\langle\bar{s}{s}\rangle}{16\pi^2}\int_{\alpha_{i}}^{\alpha_{f}}d\alpha
\delta(p_0-\widetilde{\widetilde{m}}_Q)  \nonumber \\
&&-\frac{1}{48\pi^2}\langle\frac{\alpha_sGG}{\pi}\rangle\int_{\alpha_{i}}^{\alpha_{f}}d\alpha
\int_{\beta_{i}}^{1-\alpha} d\beta (1-\alpha-\beta)
\left[p_0+\frac{p_0^2}{8}\delta(p_0-\widetilde{m}_Q)\right]\nonumber\\
&&-\frac{m_s\langle\bar{s}g_s\sigma
G{s}\rangle}{32\pi^2}\int_{\alpha_{i}}^{\alpha_{f}}d\alpha
\alpha(1-\alpha)\left[1+
\frac{2p_0}{3T}+\frac{p_0^2}{6T^2}\right]\delta(p_0-\widetilde{\widetilde{m}}_Q)\,,
\end{eqnarray}

\begin{eqnarray}
\rho^B_{\Omega^*_{QQ}}(p_0)&=&\frac{3m_s}{32 \pi^4}
\int_{\alpha_{i}}^{\alpha_{f}}d\alpha \int_{\beta_{i}}^{1-\alpha}
d\beta\alpha\beta(p_0^2-\widetilde{m}^2_Q)(3p_0^2-2\widetilde{m}^2_Q)
\nonumber\\
&&+\frac{3m_sm_Q^2}{16\pi^4}\int_{\alpha_{i}}^{\alpha_{f}}d\alpha
\int_{\beta_{i}}^{1-\alpha} d\beta
(p_0^2-\widetilde{m}^2_Q) \nonumber\\
&&-\frac{m_sm_Q^2}{384\pi^2T}
\langle\frac{\alpha_sGG}{\pi}\rangle\int_{\alpha_{i}}^{\alpha_{f}}d\alpha
\int_{\beta_{i}}^{1-\alpha} d\beta
\left[\frac{\alpha}{\beta^2}+\frac{\beta}{\alpha^2} \right] \delta(p_0-\widetilde{m}_Q)\nonumber\\
&&-\frac{m_sm_Q^4}{384\pi^2p_0^2T}\langle\frac{\alpha_sGG}{\pi}\rangle\int_{\alpha_{i}}^{\alpha_{f}}d\alpha
\int_{\beta_{i}}^{1-\alpha} d\beta \left[\frac{1}{\alpha^3}+\frac{1}{\beta^3}\right]\delta(p_0-\widetilde{m}_Q)\nonumber\\
&&+\frac{m_sm_Q^2}{64\pi^2p_0}\langle\frac{\alpha_sGG}{\pi}\rangle\int_{\alpha_{i}}^{\alpha_{f}}d\alpha
\int_{\beta_{i}}^{1-\alpha} d\beta \left[\frac{1}{\alpha^2}+\frac{1}{\beta^2}\right]\delta(p_0-\widetilde{m}_Q)\nonumber\\
&&-\frac{\langle\bar{s}{s}\rangle}{4\pi^2}\int_{\alpha_{i}}^{\alpha_{f}}d\alpha
\alpha(1-\alpha)\left[ 2p_0^2-\widetilde{\widetilde{m}}_Q^2\right]
-\frac{m_Q^2\langle\bar{s}{s}\rangle}{4\pi^2}\int_{\alpha_{i}}^{\alpha_{f}}d\alpha
 \nonumber \\
&&-\frac{m_s}{64\pi^2}\langle\frac{\alpha_sGG}{\pi}\rangle\int_{\alpha_{i}}^{\alpha_{f}}d\alpha
\int_{\beta_{i}}^{1-\alpha} d\beta
\left[1+\frac{p_0}{6}\delta(p_0-\widetilde{m}_Q) \right] \nonumber\\
&&+\frac{\langle\bar{s}g_s\sigma
G{s}\rangle}{16\pi^2}\int_{\alpha_{i}}^{\alpha_{f}}d\alpha
\alpha(1-\alpha)\left[3+\left(
2p_0+\frac{p_0^2}{2T}\right)\delta(p_0-\widetilde{\widetilde{m}}_Q)\right]\,,
\end{eqnarray}
where $\alpha_{f}=\frac{1+\sqrt{1-4m_Q^2/p_0^2}}{2}$,
$\alpha_{i}=\frac{1-\sqrt{1-4m_Q^2/p_0^2}}{2}$,
$\beta_{i}=\frac{\alpha m_Q^2}{\alpha p_0^2 -m_Q^2}$,
$\widetilde{m}_Q^2=\frac{(\alpha+\beta)m_Q^2}{\alpha\beta}$,
$\widetilde{\widetilde{m}}_Q^2=\frac{m_Q^2}{\alpha(1-\alpha)}$
   in the channels $\Xi_{QQ}({\frac{1}{2}^-})$, $\Omega_{QQ}({\frac{1}{2}^-})$,  $\Xi^*_{QQ}({\frac{3}{2}^-})$ and $\Omega^*_{QQ}({\frac{3}{2}^-})$; and $\widetilde{m}_Q^2=\frac{m_Q^2}{t}$,
$t_i=\frac{m_Q^2}{p_0^2}$     in the channels
$\Sigma_Q({\frac{1}{2}^-})$, $\Xi'_Q({\frac{1}{2}^-})$,
$\Omega_Q({\frac{1}{2}^-})$, $\Sigma_Q^*({\frac{3}{2}^-})$,
$\Xi^*_Q({\frac{3}{2}^-})$ and $\Omega^*_Q({\frac{3}{2}^-})$; the
$\rm{RP}$ denotes the replacements
$\langle\bar{s}s\rangle\rightarrow\langle\bar{q}q\rangle$,
 $\langle\bar{s}g_s\sigma G s\rangle\rightarrow\langle\bar{q}g_s\sigma
G q\rangle$ and $m_s \rightarrow 0$.

\end{document}